\documentclass{tlp}
\usepackage{amssymb}
\usepackage{latexsym}
\usepackage{verbatim}
\usepackage{color}
\usepackage[dvips]{graphicx}

\newcommand{\OP}{\mathit{OP}}

\newcommand{\da}{\downarrow}
\newcommand{\proj}[1]{\downarrow_{#1}}

\newcommand{\dqed}{\ \ \hfill $\Box$ \medskip}

\newcommand{\sep}{\,\, | \,\,}

\newcommand{\logic}[1]{\mathbf{#1}}

\newcommand{\class}[1]{\mathsf{K}}
\newcommand{\mc}[1]{\mathcal{#1}}
\newcommand{\op}[1]{\mathbf{#1}}

\newcommand{\And} {\, \wedge \,}
\newcommand{\Or} {\, \vee \,}

\newcommand{\Imp} {\, \rightarrow \,}
\newcommand{\IImp} {\, \leftrightarrow \,}
\newcommand{\A} {\forall}
\newcommand{\E} {\exists}

\newcommand{\Next}{\bigcirc \,}

\newcommand{\tl}[2]{{\mathbf{#1}(\mathbf{#2})}}

\newcommand{\ptl}{{\rm PLTL}}
\newcommand{\pltl}{{\rm PLTL}}
\newcommand{\qltl}{{\rm QLTL}}

\newcommand{\ctls}{{\rm CTL}^*}

\newcommand{\dctls}{{\rm CTL}^{*}_{\rm k}}
\newcommand{\qdctls}{{\rm QCTL}^{*}_{\rm k}}
\newcommand{\eqltl}{{\rm EQLTL}}

\newcommand{\eqdctls}{{\rm EQCTL}^{*}_{\rm k}}

\newtheorem{theorem}{Theorem}[section]
\newtheorem{lemma}[theorem]{Lemma}
\newtheorem{corollary}[theorem]{Corollary}
\newtheorem{definition}[theorem]{Definition}
\newtheorem{example}[theorem]{Example}

\newcommand{\thC}[1] {{\rm (}#1{\rm )}\nopagebreak\par\smallskip\nopagebreak}

\input epsf

\submitted{18 March 2002}

\revised{14 January 2003}

\accepted{5 September 2003}

\title[Temporalized logics and automata for time granularity]{Temporalized logics and automata
\\ for time granularity}

\author[Massimo Franceschet and Angelo Montanari]{MASSIMO FRANCESCHET\\
          Department of Sciences, University of Chieti-Pescara,
          Italy \\
          \email{francesc@sci.unich.it}
          \and ANGELO MONTANARI\\
          Department of Mathematics and Computer Science, University of Udine,
          Italy \\
          \email{montana@dimi.uniud.it}}

\date{}

\begin{document}

\DeclareGraphicsExtensions{.pstex} 

\maketitle
\begin{abstract}
The ability of providing and relating temporal representations at
different `grain levels' of the same reality is an important
research theme in computer science and a major requirement for
many applications, including formal specification and
verification, temporal databases, data mining, problem solving,
and natural language understanding. In particular, the addition of
a granularity dimension to a temporal logic makes it possible to
specify in a concise way reactive systems whose behaviour can be
naturally modeled with respect to a (possibly infinite) set of
differently-grained temporal domains.

Suitable extensions of the monadic second-order theory of $k$
successors have been proposed in the literature to capture the
notion of time granularity. In this paper, we provide the monadic
second-order theories of downward unbounded layered structures,
which are infinitely refinable structures consisting of a coarsest
domain and an infinite number of finer and finer domains, and of
upward unbounded layered structures, which consist of a finest
domain and an infinite number of coarser and coarser domains, with
expressively complete and elementarily decidable temporal logic
counterparts.

We obtain such a result in two steps. First, we define a new class
of combined automata, called temporalized automata, which can be
proved to be the automata-theoretic counterpart of temporalized
logics, and show that relevant properties, such as closure under
Boolean operations, decidability, and expressive equivalence with
respect to temporal logics, transfer from component automata to
temporalized ones. Then, we exploit the correspondence between
temporalized logics and automata to reduce the task of finding the
temporal logic counterparts of the given theories of time
granularity to the easier one of finding temporalized automata
counterparts of them.
\end{abstract}

\section{Introduction}
\label{sec:int}

Time granularity is an important, but not always well-understood,
research theme in computer science. To acquaint the reader with
the basics of the subject, we start the paper with a gentle
introduction to research on time granularity. In
Section~\ref{sec:granularity}, we briefly illustrate the
intersection of research on time granularity with different areas
of computer science, ranging from system specification and
verification to natural language understanding, and we give a
high-level view of the logical approach to the problem of
representing and reasoning about time granularity that we follow
in the paper. In Section~\ref{sec:contr}, we focus on the topics
addressed in the paper, and we outline its main contributions. In
Section~\ref{sec:reliss}, we show that the considered topics
present interesting connections with a number of issues relevant
to various research directions in computer science logic,
including real-time logics, interval logics, and combined logics.
We conclude the introduction by a short description of the
organization of the rest of the paper.

\subsection{Representing and reasoning about time granularity}
\label{sec:granularity}


The ability of providing and relating temporal representations at
different `grain levels' of the same reality is an important
research theme in various fields of computer science, including
formal specification and verification, temporal databases, data
mining, problem solving, and natural language understanding. As
for {\em formal specifications\/}, there exists a large class of
reactive systems whose components have dynamic behavior regulated
by very different time constants (granular reactive systems). A
good specification language must enable one to specify and verify
the components of a granular reactive system and their
interactions in a simple and intuitively clear
way~\cite{CCMSP93,CCMMMPR91,CMR91b,FiMa94,Lam85,MPP02,MPP99,MPP00,MP96}.
As for {\em temporal databases\/}, the common way to represent
temporal information is to timestamp either attributes
(\emph{attribute timestamping}) or tuples/objects
(\emph{tuple-timestamping}). Timestamping is performed taking time
values over some fixed granularity. However, it may happen that
differently-grained timestamps are associated with different data.
This is the case, for instance, when information is collected from
distinct sources which are not under the same control. Moreover,
users and application programs may require the flexibility of
viewing and querying temporal data at different time
granularities.  To guarantee consistency either the data must be
converted into a uniform granularity-independent representation or
temporal database operations must be generalized to cope with data
associated with different temporal domains.  In both cases, a
precise semantics for time granularity is
needed~\cite{BBJW97,CSS93,CP01,DS95,JLW93,JSW95,MP93,NJW02,NS93,SC93,Wi98,W99}.
With regard to {\em data mining}, a huge amount of data is
collected every day in the form of event-time sequences. These
sequences represent valuable sources of information, not only for
what is explicitly recorded, but also for deriving implicit
information and predicting the future behavior of the monitored
process. This latter activity requires an analysis of the
frequency of certain events, the discovery of their regularity,
and the identification of sets of events that are linked by
particular temporal relationships. Such frequencies, regularity,
and relationships are often expressed in terms of multiple
granularities, and thus analysis and discovery tools must be able
to cope with them~\cite{AS95,BJLW98,BJW96b,DDS94,MTV95}. With
regard to {\em problem solving\/}, several problems in scheduling,
planning, and diagnosis can be formulated as temporal constraint
satisfaction problems provided with a time granularity dimension.
Variables are used to represent events occurring at different time
granularities and constraints are used to represent temporal
relations between events
~\cite{BWJ96a,CD98,E95,L87,MMCR92,MR96,PB91,S96}. Finally, shifts
in the temporal perspective are common in natural language
communication, and thus the ability of supporting and relating a
variety of temporal models, at different grain sizes, is a
relevant feature for the task of {\em natural language
processing\/}~\cite{BB03,FLM86,FGMT89,KS01}.


\medskip

According to a commonly accepted perspective, any time granularity
can be viewed as the partitioning of a temporal domain in groups
of elements, where each group is perceived as an indivisible unit
(a granule). A representation formalism can then use these
granules to provide facts, actions or events with a temporal
qualification, at the appropriate abstraction level. However,
adding the concept of time granularity to a formalism does not
merely mean that one can use different temporal units to represent
temporal quantities in a unique flat model, but it involves
semantic issues related to the problem of assigning a proper
meaning to the association of statements with the different
temporal domains of a layered model and of switching from one
domain to a coarser/finer one.

\begin{figure}
\centering
\input{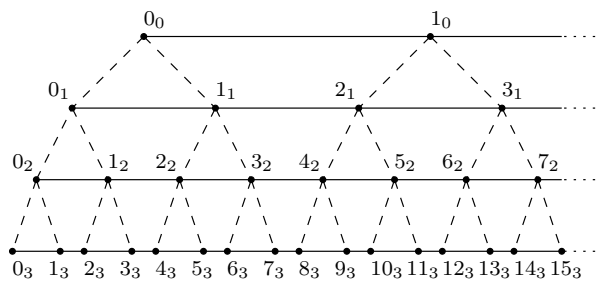} 
\caption{\label{fig:fin} The $2$-refinable 4-layered structure.}
\end{figure}

Different approaches to represent and to reason about time
granularity have been proposed in the literature. In the
following, we introduce the distinctive features of the logical
approach to time granularity\footnote{In \cite{FM02b} we analyze
alternative approaches to time granularity, developed in the
context of temporal databases, and we compare them with the
logical one.}. In the {\em logical} setting, the different time
granularities and their interconnections are represented by means
of mathematical structures, called layered structures. A {\em
layered structure} consists of a possibly infinite set of related
differently-grained temporal domains. Such a structure identifies
the relevant temporal domains and defines the relations between
time points belonging to different domains. Suitable operators
make it possible to move horizontally {\em within} a given
temporal domain (displacement operators), and to move vertically
{\em across} temporal domains (projection operators). Both
classical and temporal logics can be interpreted over the layered
structure. Logical formulas allow one to specify properties
involving different time granularities in a single formula by
mixing displacement and projection operators. Algorithms are
provided to verify whether a given formula is consistent
(\emph{satisfiability checking}) as well as to check whether a
given formula is satisfied in a particular structure (\emph{model
checking}). The logical approach to represent time granularity has
been mostly applied in the field of formal specification and
verification of concurrent systems. An application of time
granularity logics to the specification of a supervisor that
automates the activities of a high voltage station, devoted to the
end user distribution of the energy generated by power plants, has
been accomplished in collaboration with Automation Research Center
of the Electricity Board of Italy (ENEL). A short account of this
work has been given in \cite{CCMSP93}. Logics for time granularity
have also been applied to the specification of real-time
monitoring systems \cite{CCMMMPR91}, mobile systems
\cite{FMdeR00}, and therapy plans in clinical medicine
\cite{CFP02}.

\begin{figure}
\centering
\input{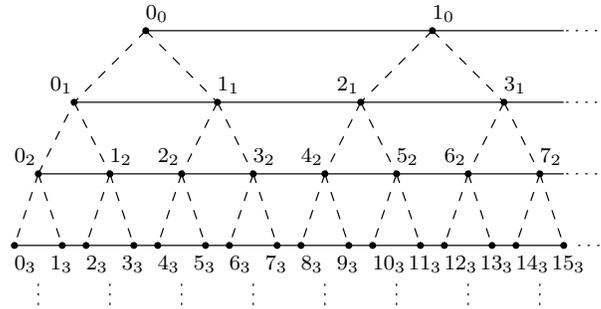} 
\caption{\label{fig:down} The $2$-refinable downward unbounded
layered structure.}
\end{figure}

A systematic logical framework for time granularity, based on a
many-level view of temporal structures, with matching logics and
decidability results, has been proposed in~\cite{M96,MP96,MPP99}
and later extended in~\cite{F01,FM01b,FM01a,FM03}. Layered
structures with exactly $n \geq 1$ temporal domains such that each
time point can be refined into $k \geq 2$ time points of the
immediately finer temporal domain, if any, are called
$k$-refinable $n$-layered structures ($n$-LSs for short, see
Figure~\ref{fig:fin}). They have been investigated in~\cite{MP96},
where a classical second-order language, with second-order
quantification restricted to monadic predicates, has been
interpreted over them. The language includes a total order $<$ and
$k$ projection functions ${\downarrow}_0, \ldots,
{\downarrow}_{k-1}$ over the layered temporal universe such that,
for every point $x$, ${\downarrow}_0(x), \ldots, {\downarrow}_{k -
1}(x)$ are the $k$ elements of the immediately finer temporal
domain, if any, into which $x$ is refined. The satisfiability
problem for the monadic second-order language over $n$-LSs has
been proved to be decidable by using a reduction to the emptiness
problem for B\"uchi sequence automata. Unfortunately, the decision
procedure has a nonelementary complexity.

\begin{figure}
\centering
\input{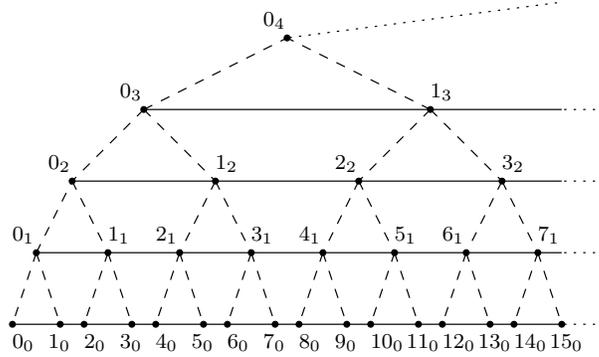} 
\caption{\label{fig:up} The $2$-refinable upward unbounded layered
structure.}
\end{figure}

Layered structures with an infinite number of temporal domains,
$\omega$-layered structures, have been studied in~\cite{MPP99}. In
particular, the authors investigated {\em $k$-refinable downward
unbounded layered structures} (DULSs), that is, $\omega$-layered
structures consisting of a coarsest domain together with an
infinite number of finer and finer domains (see
Figure~\ref{fig:down}), and {\em $k$-refinable upward unbounded
layered structures} (UULSs), that is, $\omega$-layered structures
consisting of a finest temporal domain together with an infinite
number of coarser and coarser domains (see Figure~\ref{fig:up}). A
classical monadic second-order language, including a total order
$<$ and $k$ projection functions ${\downarrow}_0, \ldots,
{\downarrow}_{k-1}$, has been interpreted over both UULSs and
DULSs. The decidability of the monadic second-order theories of
UULSs and DULSs has been proved by reducing the satisfiability
problem to the emptiness problem for systolic and Rabin tree
automata, respectively. In both cases the decision procedure has a
nonelementary complexity.

\subsection{Our contributions}
\label{sec:contr}

Monadic logics for time granularity are quite expressive, but,
unfortunately, they have few computational appealing: their
decision problem is indeed \emph{nonelementary}. This roughly
means that it is possible to algorithmically check satisfiability,
but the complexity of the algorithm grows very rapidly and cannot
be bounded. Moreover, the corresponding automata (B\"uchi sequence
automata for the theory of finitely-layered structures, Rabin tree
automata for downward unbounded structures, and systolic tree
automata for upward unbounded ones) do not directly work over
layered structures, but rather over collapsed structures into
which layered structures can be encoded. Hence, they are not
natural and intuitive tools to specify and check properties of
time granularity.

In this paper, we follow a different approach. Taking inspiration
from combination methods for temporal logics, we start by studying
how to combine automata in such a way that properties of the
components are inherited by the combination. Then, we reinterpret
layered structures as \emph{combined structures}. This intuition
reveals to be the keystone of our endeavor. Indeed, it allows us
to define combined temporal logics and combined automata over
layered structures, and to study their expressive power and
computational properties by taking advantage of the transfer
theorems for combined logics and combined automata. The outcome is
appealing: the resulting combined temporal logics and automata
directly work over layered structures. Moreover, they are
expressively equivalent to monadic languages, and they are
elementarily decidable.

\begin{figure}[t]
\begin{center}
\begin{picture}(150,150)
\put(-35,90){{\bf Monadic}} \put(-35,80){{\bf Theories}}
\put(100,90){{\bf (Temporalized)}} \put(120,80){{\bf Logics}}
\put(15,16){{\bf (Temporalized)}} \put(25,6){{\bf Automata}}
\put(15,85){\vector(1,0){80}} \put(95,85){\vector(-1,0){80}}
\put(5,75){\vector(1,-1){45}} \put(5,75){\vector(-1,1){1}}
\put(100,75){\vector(-1,-1){45}} \put(100,75){\vector(1,1){1}}
\put(55,95){?}
\end{picture}
\caption{\label{fig:dia} From monadic theories to (temporalized)
logics via (temporalized) automata.}
\end{center}
\end{figure}
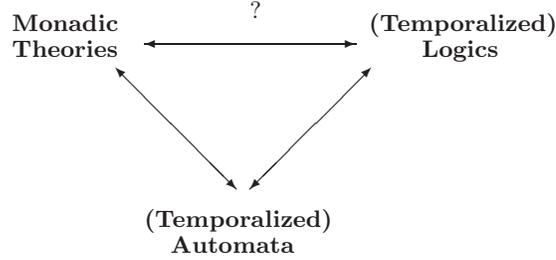

Finding the temporal logic counterpart of monadic theories is a
difficult task, involving a nonelementary blow up in the length of
formulas. Ehrenfeucht games have been successfully exploited to
deal with such a correspondence problem for first-order monadic
theories~\cite{IK89} and well-behaved fragments of second-order
ones, e.g. the path fragment of the monadic second-order theory of
infinite binary trees~\cite{HT87}. As for the theories of time
granularity, by means of suitable applications of Ehrenfeucht
games, we obtained an expressively complete and elementarily
decidable combined temporal logic counterpart of the path fragment
of the monadic second-order theory of DULSs~\cite{FM03}, while
Montanari et al. extended Kamp's theorem to deal with the
first-order fragment of the theory of UULSs~\cite{MPP02}.
Unfortunately, these techniques produce rather involved proofs and
do not naturally lift to the full second-order case.

In this paper, instead of trying to establish a direct
correspondence between monadic second-order theories for time
granularity and temporal logics, we connect them via automata
(cf.\ Figure~\ref{fig:dia}). Firstly, we define a new class of
combined automata, called temporalized automata, which can be
proved to be the automata-theoretic counterpart of temporalized
logics, and show that relevant properties, such as closure under
Boolean operations, decidability, and expressive equivalence with
respect to temporal logics, transfer from component automata to
temporalized ones. Then, on the basis of the established
correspondence between temporalized logics and automata, we reduce
the task of finding a temporal logic counterpart of the monadic
second-order theories of DULSs and UULSs to the easier one of
finding a temporalized automata counterpart of them. The mapping
of monadic formulas into automata (the difficult direction) can
indeed greatly benefit from automata closure properties.

As a by-product, the alternative characterization of temporalized
logics for time granularity as temporalized automata allows one to
reduce logical problems to automata ones. As it is well-known in
the area of automated system specification and verification, such
a reduction presents several advantages, including the possibility
of using automata for both system modeling and specification, and
the possibility of checking the system on-the-fly (a detailed
account of these advantages can be found in~\cite{FM01a}).

\subsection{Related fields}
\label{sec:reliss}

The original motivation of our research was the design of a
temporal logic embedding the notion of time granularity, suitable
for the specification of complex concurrent systems whose
components evolve according to different time units. However, we
soon established a fruitful complementary point of view on time
granularity: it can be regarded as a powerful setting to
investigate the definability of meaningful timing properties over
a \emph{single} time domain. Moreover, layered structures and
logics provide an interesting embedding framework for flat
\emph{real-time} structures and logics, as well as there exists a
natural link between structures and theories of time granularity
and those developed for representing and reasoning about
\emph{time intervals}. Finally, there are significant similarities
between the problems we encountered in studying time granularity,
and those addressed by current research on \emph{combining}
logics, theories, and structures. In the following, we briefly
explain all these connections.

\paragraph{Granular reactive systems.}
As pointed out above, we were originally motivated by the design
of a temporal logic embedding the notion of time granularity
suitable for the specification of granular reactive systems. A
{\em reactive system} is a concurrent program that maintains and
interaction with the external environment and that ideally runs
forever. Temporal logic has been successfully used for modeling
and analyzing the behavior of reactive systems \cite{E90}. It
supports semantic model checking, which can be used to check
specifications against system behaviors; it also supports pure
syntactic deduction, which may be used to verify the consistency
of specifications. Finite-state automata, such as B\"uchi sequence
automata and Rabin tree automata \cite{T90}, have been proved very
useful in order to provide clean and asymptotically optimal
satisfiability and model checking algorithms for temporal
logics~\cite{KVW00,VW94} as well as to cope with the {\em state
explosion problem} that frightens concurrent system
verification~\cite{CVWY91,JJ89,VW86}.

A {\em granular reactive systems} is a reactive system whose
components have dynamic behaviours regulated by very different
time constants~\cite{M96}. As an example, consider a pondage power
station consisting of a reservoir, with filling and emptying times
of days or weeks, generator units, possibly changing state in a
few seconds, and electronic control devices, evolving in
microseconds or even less. A complete specification of the power
station must include the description of these components and of
their interactions. A natural description of the temporal
evolution of the reservoir state will probably use days: ``During
rainy weeks, the level of the reservoir increases 1 meter a day'',
while the description of the control devices behaviour may use
microseconds: ``When an alarm comes from the level sensors, send
an acknowledge signal in 50 microseconds''. We say that systems of
such a type have {\em different time granularities}. It is
somewhat unnatural, and sometimes impossible, to compel the
specifier to use a unique time granularity, microseconds in the
previous example, to describe the behaviour of all the components.
A good language must indeed allow the specifier to easily describe
all simple and intuitively clear facts (naturalness of the
notation). Hence, a specification language for granular reactive
systems must support different time granularities to allow one (i)
to maintain the specifications of the dynamics of
differently-grained components as separate as possible (modular
specifications), (ii) to differentiate the refinement degree of
the specifications of different system components (flexible
specifications), and (iii) to write complex specifications in an
incremental way by refining higher-level predicates associated
with a given time granularity in terms of more detailed ones at a
finer granularity (incremental specifications).

\paragraph{Definability of meaningful timing properties.}
Time granularity can be viewed not only as an important feature of
a representation language, but also as a formal tool to
investigate the definability of meaningful timing properties, such
as density and exponential grow/decay, over a {\em single} time
domain~\cite{MPP99}. In this respect, the number of layers (single
vs.\ multiple, finite vs.\ infinite) of the underlying temporal
structure, as well as the nature of their interconnections, play a
major role: certain timing properties can be expressed using a
single layer; others using a finite number of layers; others only
exploiting an infinite number of layers. For instance, temporal
logics over binary $2$-layered structures suffice to deal with
conditions like ``$P$ holds at all even times of a given temporal
domain'' that cannot be expressed using flat propositional
temporal logics~\cite{W83}. Moreover, temporal logics over
$\omega$-layered structures allow one to express relevant
properties of infinite sequences of states over a single temporal
domain that cannot be captured by using flat or $n$-layered
temporal logics. For instance, temporal logics over $k$-refinable
UULSs allow one to express conditions like ``$P$ holds at all time
points $k^i$, for all natural numbers $i$, of a given temporal
domain'', which cannot be expressed by using  either propositional
or quantified temporal logics over a finite number of layers,
while temporal logics over DULSs allow one to constrain a given
property to hold true `densely' over a given time interval.

\paragraph{On the relationship with real-time logics.}
Layered structures and logics can be regarded as an embedding
framework for flat real-time structures and logics. A {\em
real-time system} is a reactive system with well-defined
fixed-time constraints. Systems that control scientific
experiments, industrial control systems, automobile-engine
fuel-injection systems, and weapon systems are examples of
real-time systems. Examples of quantitative timing properties
relevant to real-time systems are periodicity, bounded
responsiveness, and timing delays. Logics for real-time systems,
called real-time logics, are interpreted over {\em timed state
sequences}, that is, state sequences in which every state is
associated with a time instant.

Montanari et al.\ showed that the second-order theory of timed
state sequences can be {\em properly} embedded into the
second-order theory of binary UULSs as well as into the
second-order theory of binary DULSs~\cite{MPP00}. The increase in
expressive power of the embedding frameworks makes it possible to
express and check additional timing properties of real-time
systems, which cannot be dealt with by the classical theory. For
instance, in the theory of timed state sequences, saying that a
state $s$ holds true at time $i$ can be meant to be an abstraction
of the fact that state $s$ can be arbitrarily placed in the time
interval $[i,i+1)$.  The stratification of domains in layered
structures naturally supports such an interval interpretation and
gives means for reducing the uncertainty involved in the
abstraction process. For instance, it allows on to say that a
state $s$ belongs to the first (respectively, second) half of the
time interval $[i,i+1)$. More generally, the embedding of
real-time logics into the granularity framework allows one to deal
with {\em temporal indistinguishability} of states (two or more
states associated with the same time) and {\em temporal gaps}
between states (a nonempty time interval between the time
associated to two contiguous states). Temporal
indistinguishability and temporal gaps can indeed be interpreted
as phenomena due to the fact that real-time logics lack the
ability to express properties at the right (finer) level of
granularity: distinct states, associated with the same time, can
always be ordered at the right level of granularity; similarly,
time gaps represent intervals in which a state cannot be specified
at a finer level of granularity. A finite number of layers is
obviously not sufficient to capture timed state sequences: it is
not possible to fix a priori any bound on the granularity that a
domain must have to allow one to temporally order a given set of
states, and thus we need to have an infinite number of temporal
domains at our disposal.

\paragraph{On the relationship with interval logics.}
As pointed out in~\cite{M96}, there exists a natural link between
structures and theories of time granularity and those developed
for representing and reasoning about time intervals.
Differently-grained temporal domains can indeed be interpreted as
different ways of partitioning a given discrete/dense time axis
into consecutive disjoint intervals. According to this
interpretation, every time point can be viewed as a suitable
interval over the time axis and projection implements an
intervals-subintervals mapping. More precisely, let us define {\em
direct constituents} of a time point $x$, belonging to a given
domain, the time points of the immediately finer domain into which
$x$ can be refined, if any, and {\em indirect constituents} the
time points into which the direct constituents of $x$ can be
directly or indirectly refined, if any. The mapping of a given
time point into its direct or indirect constituents can be viewed
as a mapping of a given time interval into (a specific subset of)
its subintervals.

The existence of such a natural correspondence between interval
and granularity structures hints at the possibility of defining a
similar connection at the level of the corresponding theories. For
instance, according to such a connection, temporal logics over
DULSs allow one to constrain a given property to hold true densely
over a given time interval, where $P$ densely holds over a time
interval $w$ if $P$ holds over $w$ and there exists a direct
constituent of $w$ over which $P$ densely holds. In particular,
establishing a connection between structures and logics for time
granularity and those for time intervals would allow one to
transfer decidability results from the granularity setting to the
interval one. As a matter of fact, most interval temporal logics,
including Moszkowski's Interval Temporal Logic (ITL)~\cite{M83},
Halpern and Shoham's Modal Logic of Time Intervals
(HS)~\cite{HS91}, Venema's CDT Logic~\cite{V91}, and Chaochen and
Hansen's Neighborhood Logic (NL)~\cite{CH98}, are highly
undecidable. Decidable fragments of these logics have been
obtained by imposing severe restrictions on their expressive
power, e.g., the {\em locality} constraint in \cite{M83}.

Preliminary results can be found in~\cite{MSV02}, where the
authors propose a new interval temporal logic, called Split Logic
(SL for short), which is equipped with operators borrowed from HS
and CDT, but is interpreted over specific interval structures,
called {\em split-frames\/}. The distinctive feature of a
split-frame is that there is at most one way to chop an interval
into two adjacent subintervals, and consequently it does not
possess {\em all} the intervals. They prove the decidability of SL
with respect to particular classes of split-frames which can be
put in correspondence with the first-order fragments of the
monadic theories of time granularity. In particular, {\em
discrete} split-frames with maximal intervals correspond to
finitely layered structures, discrete split-frames (with unbounded
intervals) can be mapped into upward unbounded layered structures,
and {\em dense} split-frames with maximal intervals can be encoded
into downward unbounded layered structures.

\paragraph{The combining logic perspective.}
There are significant similarities between the problems we
addressed in the time granularity setting and those dealt with by
current research on logics that model changing contexts and
perspectives. The design of these types of logics is emerging as a
relevant research topic in the broader area of combination of
logics, theories, and structures, at the intersection of logic
with artificial intelligence, computer science, and computational
linguistics~\cite{GdR00}. The reason is that application domains
often require rather complex hybrid description and specification
languages, while theoretical results and implementable algorithms
are at hand only for simple basic components~\cite{GKWZ03}. As for
granular reactive systems, their operational behavior can be
naturally described as a suitable combination of temporal {\em
evolutions} (sequences of component states) and temporal {\em
refinements} (mapping of a component state into a finite sequence
of states belonging to a finer component). According to such a
point of view, the model describing the operational behavior of
the system and the specification language can be obtained by {\em
combining} simpler models and languages, respectively, and model
checking/satisfiability procedures for combined logics can be
used.

\bigskip \noindent
From the above discussion, it turns out that the time granularity
framework is expressive and flexible enough to be used to
investigate many interesting topics not explicitly related to time
granularity. The aim of this paper is to deepen our understanding
of time granularity. The rest of the paper is organized as
follows. In Section \ref{sec:tempaut}, we introduce temporalized
automata and we show that relevant logical properties, such as
closure under Boolean operations and decidability, transfer from
component automata to temporalized ones; furthermore, we prove
that temporalized automata are as expressive as temporalized
logics. In Section \ref{sec:timegran} we exploit temporalized
automata to find the temporal logic counterparts of the given
theories of time granularity. Temporalized automata for the
theories of DULSs and UULSs are obtained as combinations of
B\"uchi and Rabin automata and of B\"uchi and finite tree
automata, respectively. As a matter of fact, unlike the case of
DULSs, the combined model we use to encode an UULS differs from
that of pure temporalization since the innermost submodels are not
independent from the outermost top-level model. In Section
\ref{sec:exa}, we apply temporalized logics to a real-world case
study. Conclusive remarks provide an assessment of the work done
and outline some future research directions.

\section{Temporalized logics and automata}
\label{sec:tempaut}

In this section we recall the definition of temporalization and we
define temporalized automata\footnote{We assume the reader to be
familiar with basic concepts of modal and temporal logics, and
automata. If this is not the case, comprehensive surveys are given
in~\cite{E90} and~\cite{T90}, respectively.}. Moreover, we prove
the equivalence of temporalized automata and temporalized logics.
We will take into consideration the following well-known temporal
logics: {\em Propositional Linear Temporal Logic} ($\pltl$), {\em
Quantified Linear Temporal Logic} ($\qltl$), {\em Existentially
Quantified Linear Temporal Logic} ($\eqltl$), {\em Directed
Computational Tree Logic} ($\dctls$), {\em Quantified Directed
Computational Tree Logic} ($\qdctls$), and {\em Existentially
Quantified Directed Computational Tree Logic} ($\eqdctls$);
moreover, we will take advantage of the following well-known
finite-state automata classes: B\"uchi sequence automata, Rabin
tree automata, finite tree automata.

Let ${\cal P} = \{P,Q, \ldots\}$ be a set of proposition letters.
We consider {\em temporal logics} over the set of propositional
letters ${\cal P}$. Given a temporal logic $\logic{T}$, we use
${\cal L}_{\logic{T}}$ and $\mathsf{K}_{\logic{T}}$ to denote the
language and the set of models of $\logic{T}$, respectively.
Furthermore, we write $\mathit{OP}(\logic{T})$ to denote the set
of temporal operators of $\logic{T}$.

{\em Temporalization} is a simple form of logic combination that
embeds one component logic into the other~\cite{FG92}. Let
$\logic{T}$ be a temporal logic and $\logic{L}$ an arbitrary
logic. For the sake of simplicity, we constrain $\logic{L}$ to be
an extension of propositional logic. We partition the set of
$\logic{L}$-formulas into {\em Boolean combinations}
$\mathit{BC}_{\logic{L}}$ and {\em monolithic formulas}
$\mathit{ML}_{\logic{L}}$: $\alpha$ belongs to
$\mathit{BC}_{\logic{L}}$ if its outermost operator is a Boolean
connective; otherwise it belongs to $\mathit{ML}_{\logic{L}}$.  We
assume that $\OP(\logic{T}) \cap \OP(\logic{L}) = \emptyset$.

\begin{definition} \thC{Temporalization -- Syntax}
\label{def:colog1} \noindent The language ${\cal L}_{\tl{T}{L}}$
of the \emph{temporalization} $\tl{T}{L}$ of $\logic{L}$ by means
of $\logic{T}$ over the set of proposition letters $\cal P$ is
obtained by taking the set of formation rules of ${\cal
L}_{\logic{T}}$ and by replacing the atomic formation rule:
``every proposition letter $P \in {\cal P}$ is a formula'' by the
rule: ``every monolithic formula $\alpha \in
\mathcal{L}_{\logic{L}}$ is a formula''. \dqed
\end{definition}

\noindent As an example, let $\logic{T_1}$ and $\logic{T_2}$ be
two temporal logics, and let $\{\op{F_1}, \op{G_1}\}$ (resp.
$\{\op{F_2}, \op{G_2}\}$) be the temporal operators of
$\logic{T_1}$ (resp. $\logic{T_2}$). The formula ${\bf F_1 G_2} p$
is a $\tl{T_1}{T_2}$-formula, while the formula ${\bf F_1 G_2} p
\IImp {\bf G_2 F_1} p$ is not.

\medskip

A {\em model} for $\tl{T}{L}$ is a triple $(W,{\cal R},g)$, where
$(W,{\cal R})$ is a frame for $\logic{T}$ and $g \, : \, W \Imp
\mathsf{K}_{\logic{L}}$ a total function mapping worlds in $W$ to
models for $\logic{L}$.

\begin{definition} \thC{Temporalization -- Semantics}
\label{def:cosem1} \noindent Given a model ${\cal M} = (W,{\cal
R},g)$ and a state $w \in W$, the semantics of the temporalized
logic $\tl{T}{L}$ is obtained by taking the set of semantic
clauses of $\logic{T}$ and by replacing the clause for proposition
letters: ``${\cal M}, w \models P$ if and only if $P \in V(w)$,
whenever $P \in {\cal P}$'' by the clause: ``${\cal M}, w \models
\alpha$ if and only if $g(w) \models_{\logic{L}} \alpha$, whenever
$\alpha \in ML_{\logic{L}}$''. \dqed
\end{definition}

Hereafter, we will restrict our attention to temporalized logics
such that both the embedding and the embedded logics are temporal
logics.

\medskip

We now introduce a new class of combined automata, called
\emph{temporalized automata}, which can be viewed as the
automata-theoretic counterpart of temporalized logics, and show
that relevant properties, such as closure under Boolean
operations, decidability, and expressive equivalence with respect
to temporal logics, transfer from component automata to
temporalized ones. We first define automata and prove results over
sequence structures; then, we generalize definitions and results
to tree structures (as a matter of fact, we believe that our
machinery can actually be extended to cope with more general
structures, such as graphs). We will use the following general
definition of sequence automata. Let $\Sigma = \{a,b,\ldots\}$ be
a finite alphabet and let ${\cal S}(\Sigma)$ be the set of
$\Sigma$-labeled infinite sequences, that is, structures of the
form $(\mathbb{N},<,V)$, where $(\mathbb{N},<)$ is the set of
natural numbers, together with the usual ordering relation, and $V
\, : \, \mathbb{N} \Imp \Sigma$ is a valuation function mapping
natural numbers into symbols in $\Sigma$.

\begin{definition} \thC{Sequence automata} \label{def:SA}
\noindent A {\em sequence automaton} $A$ over $\Sigma$ consists of
(i) a Labeled Transition System $(Q,q_0,\Delta,$ $M,\Omega)$,
where $Q$ is a finite set of states, $q_0 \in Q$ is the initial
state, $\Delta \subseteq Q \times \Sigma \times Q$ is a transition
relation, $\Omega$ is a finite alphabet, and $M \subseteq Q \times
\Omega$ is a labeling of states, and (ii) an acceptance condition
$AC$. Given a $\Sigma$-labeled infinite sequence $w =
(\mathbb{N},<,V)$, a {\em run} of $A$ on $w$ is a function $\sigma
\, : \, \mathbb{N} \Imp Q$ such that $\sigma(0) = q_0$ and
$(\sigma(i),V(i), \sigma(i+1)) \in \Delta$, for every $i \geq 0$.
The automaton $A$ accepts $w$ if there is a run $\sigma$ of $A$ on
$w$ such that $AC(\sigma)$, i.e., the acceptance condition holds
on $\sigma$. The language accepted by $A$, denoted by ${\cal
L}(A)$, is the set of $\Sigma$-labeled infinite sequences accepted
by $A$.~\dqed
\end{definition}

A class of sequence automata $\cal A$ is a set of automata that
share the acceptance condition $AC$ (we do not explicitly specify
the acceptance condition for sequence automata since, as we will
see, all the achieved results do not rest on any specific
acceptance condition). An example of a class of sequence automata
is the class of B\"uchi automata.

\begin{example} \thC{B\"uchi automata} \label{exa:fsa}
\noindent A B\"uchi automaton is a sequence automaton $A = (Q,
q_0, \Delta, M, \Omega)$ such that $\Omega = \{{\tt final}\}$. We
call \emph{final} a state $q$ such that $(q,{\tt final}) \in M$.
The acceptance condition for $A$ states that $A$ accepts a
$\Sigma$-labeled infinite sequence $w$ if and only if there is a
run $\sigma$ of $A$ on $w$ such that some final state occurs
infinitely often in $\sigma$. \dqed
\end{example}


Temporalized automata over sequence structures can be defined as
follows. Let ${\cal A}_2$ be a class of sequence automata which
accept sequences in ${\cal S}(\Sigma)$; moreover, let
$\Gamma(\Sigma)$ be a finite alphabet whose symbols $A,B, \ldots$
denote automata in ${\cal A}_2$, and let ${\cal A}_1$ be a class
of sequence automata which accept ($\Gamma(\Sigma)$-labeled
infinite) sequences in ${\cal S}(\Gamma(\Sigma))$. Given ${\cal
A}_1$ and ${\cal A}_2$ as above, we define a class of temporalized
automata ${\cal A}_1({\cal A}_2)$ that combine the two component
classes of automata in a suitable way. Let ${\cal S}({\cal
S}(\Sigma))$ be the set of infinite sequences of $\Sigma$-labeled
infinite sequences, that is, temporalized models
$(\mathbb{N},<,g)$ where $g \, : \, \mathbb{N} \Imp {\cal
S}(\Sigma)$ is a total function mapping elements of $\mathbb{N}$
into sequences in ${\cal S}(\Sigma)$. Automata in ${\cal
A}_1({\cal A}_2)$ accept objects in ${\cal S}({\cal S}(\Sigma))$.
The class of temporalized automata ${\cal A}_1({\cal A}_2)$ is
formally defined as follows.

\begin{definition} \thC{Temporalized automata} \label{def:combaut}
\noindent A {\em temporalized automaton} $A$ over $\Gamma(\Sigma)$
is a quintuple $(Q,q_0,\Delta, M,\Omega)$ as for sequence automata
(Definition~\ref{def:SA}). The {\em combined acceptance condition}
for $A$ is defined as follows. Given $w = (\mathbb{N},<,g) \in
{\cal S}({\cal S}(\Sigma))$, a {\em run} of $A$ on $w$ is function
$\sigma \, : \, \mathbb{N} \Imp Q$ such that $\sigma(0) = q_0$
and, for every $i \geq 0$, $(\sigma(i),B,\sigma(i+1)) \in \Delta$
for some $B \in \Gamma(\Sigma)$ such that $g(i) \in {\cal L}(B)$.
The automaton $A$ accepts $w$ if there exists a run $\sigma$ of
$A$ on $w$ such that $AC(\sigma)$, where $AC$ is the acceptance
condition of ${\cal A}_1$. The language recognized by $A$, denoted
by ${\cal L}(A)$, is the set of elements in ${\cal S}({\cal
S}(\Sigma))$ accepted by $A$.
\dqed
\end{definition}

Given a temporalized automaton $A \in {\cal A}_1({\cal A}_2)$, we
denote by $A^\uparrow$ the automaton in ${\cal A}_1$ with the same
labeling transition system as $A$ and with the acceptance
condition of ${\cal A}_1$. While $A$ accepts in ${\cal S}({\cal
S}(\Sigma))$, its {\em abstraction} $A^\uparrow$ recognizes in
${\cal S}(\Gamma(\Sigma))$.  Moreover, given an automaton $A \in
{\cal A}_1$, we denote by $A^\downarrow$ the automaton in ${\cal
A}_1({\cal A}_2)$ with the same labeling transition system  as $A$
and with the combined acceptance condition of ${\cal A}_1({\cal
A}_2)$. While $A$ accepts in ${\cal S}(\Gamma(\Sigma))$, its {\em
concretization} $A^\downarrow$ recognizes in ${\cal S}({\cal
S}(\Sigma))$. Taking advantage of these notions, the combined
acceptance condition for temporalized automata can be rewritten as
follows. Let $w = (\mathbb{N},<,g) \in {\cal S}({\cal
S}(\Sigma))$. A temporalized automaton $A$ accepts $w$ if and only
if there exists $v = (\mathbb{N},<,V) \in {\cal
S}(\Gamma(\Sigma))$ such that $v \in {\cal L}(A^\uparrow)$ and,
for every $i \in \mathbb{N}$, $g(i) \in {\cal L}(V(i))$. In the
following, we will often use this alternative, but equivalent,
formulation of the combined acceptance condition for temporalized
automata.

\medskip

We now show that relevant logical properties transfer from
component automata to temporalized ones. The following notation
will be used to express the relationships between automata and
temporal logics. We write ${\cal A} \rightarrow \logic{T}$ to
denote the fact that every automaton $A$ in $\cal A$ can be
converted into a formula $\varphi_A$ in $\logic{T}$ such that
${\cal L}(A) = {\cal M}(\varphi_A)$, where ${\cal M}(\varphi_A)$
is the set of models of $\varphi_A$. Conversely, we write
$\logic{T} \rightarrow {\cal A}$ to denote the fact that every
formula $\varphi$ in $\logic{T}$ can be converted into an
equivalent automaton in $\cal A$. Finally, ${\cal A}
\leftrightarrows \logic{T}$ stands for ${\cal A} \rightarrow
\logic{T}$ and $\logic{T} \rightarrow {\cal A}$. The {\em transfer
problem} for temporalized automata can be stated as follows.
Assuming that the automata classes ${\cal A}_1$ and ${\cal A}_2$
enjoy a given logical property, does ${\cal A}_1( {\cal A}_2)$
enjoy that property?  We investigate the transfer problem with
respect to the following properties of automata:

\begin{enumerate}
\item
(Effective) {\em closure} under Boolean operations (union,
intersection, and complementation): if ${\cal A}_1$ and ${\cal
A}_2$ are (effectively) closed under Boolean operations, is ${\cal
A}_1({\cal A}_2)$ (effectively) closed under Boolean operations?
\item
{\em Decidability}: if ${\cal A}_1$ and ${\cal A}_2$ are
decidable, is ${\cal A}_1({\cal A}_2)$ decidable?
\item
{\em Expressive equivalence} with respect to temporal logic: if
${\cal A}_1 \leftrightarrows \logic{T_1}$ and ${\cal
A}_2\leftrightarrows\logic{T_2}$, does ${\cal A}_1( {\cal
A}_2)\leftrightarrows\tl{\logic{T_1}}{\logic{T_2}}$?
\end{enumerate}

The following lemma plays a crucial role. It shows that every
temporalized automaton is equivalent to a temporalized automaton
whose transitions are labeled with automata that form a partition
of the set ${\cal S}(\Sigma)$ of $\Sigma$-labeled sequences.
Hence, {\em different} labels of the `partitioned automaton'
correspond to (automata accepting) {\em disjoint} sets of
$\Sigma$-labeled sequences. Moreover, the partitioned automaton
can be effectively constructed from the original one. We will see
that a similar partition lemma holds for temporalized logics (cf.\
Lemma~\ref{lm:part2} below).

\begin{lemma} \thC{Partition lemma for temporalized automata}
\label{lm:part} \noindent Let $A$ be a temporalized automaton in
${\cal A}_1({\cal A}_2)$. If ${\cal A}_2$ is closed under Boolean
operations (union, intersection, and complementation), then there
exists a finite alphabet $\Gamma'(\Sigma) \subseteq {\cal A}_2$
and a temporalized automaton $A'$ over $\Gamma'(\Sigma)$ such that
${\cal L}(A)={\cal L}(A')$ and the set $\{{\cal L}(X) \sep X \in
\Gamma'(\Sigma)\}$ is a partition of ${\cal S}(\Sigma)$. Moreover,
if ${\cal A}_2$ is effectively closed under Boolean operations and
it is decidable, then $A'$ can be effectively computed from $A$.
\end{lemma}

\begin{proof}
To construct $\Gamma'(\Sigma)$ and $A'$ we proceed as follows. Let
$A = (Q,q_0,\Delta,M,\Omega)$ be a temporalized automaton over
$\Gamma(\Sigma) = \{X_1, \ldots X_n\} \subseteq {\cal A}_2$ . For
every $1 \leq i \leq n$ and $j \in \{0,1\}$, let $X_{i}^{j} = X_i$
for $j=0$ and $X_{i}^{j} = {\cal S}(\Sigma) \setminus X_i$ for
$j=1$. Given $(j_1, \ldots, j_n) \in \{0,1\}^n$, let ${\tt
Cap}_{(j_1, \ldots, j_n)} = \bigcap_{i=1}^{n} X_{i}^{j_i}$. We
define $\Gamma_{1}(\Sigma)$ as the set of all and only ${\tt
Cap}_{(j_1, \ldots, j_n)}$ such that $(j_1, \ldots, j_n) \in
\{0,1\}^n$. Since ${\cal A}_2$ is closed under Boolean operations,
$\Gamma_{1}(\Sigma) \subseteq {\cal A}_2$. Moreover, let
$\Gamma_{2}(\Sigma) = \{X \in \Gamma_{1}(\Sigma) \sep {\cal L}(X)
\neq \emptyset\}$. We set $\Gamma'(\Sigma) = \Gamma_{2}(\Sigma)$,
and, for $1 \leq i \leq n$, $\Gamma_i'(\Sigma) = \{X \in
\Gamma'(\Sigma) \sep X \cap X_i \neq \emptyset\}$. Note that
$\{{\cal L}(X) \sep X \in \Gamma'(\Sigma)\}$ is a partition of
${\cal S}(\Sigma)$. Moreover, for every $1 \leq i \leq n$,
$\{{\cal L}(X) \sep X \in \Gamma_i'(\Sigma)\}$ is a partition of
${\cal L}(X_i)$. We define the temporalized automaton $A' =
(Q,q_0,\Delta',M,\Omega)$ over $\Gamma'(\Sigma)$, where $\Delta'$
contains all and only the triples $(q_1,X,q_2) \in Q \times
\Gamma'(\Sigma) \times Q$ such that $X \in \Gamma_i'(\Sigma)$ and
$(q_1,X_i,q_2) \in \Delta$ for some $1 \leq i \leq n$. It is not
difficult to see that ${\cal L}(A)={\cal L}(A')$.
\end{proof}

We now prove the first transfer theorem: closure under Boolean
operations transfers from component automata to temporalized ones.

\begin{theorem} \thC{Transfer of closure under Boolean operations}
\label{th:traclo} \noindent Closure under Boolean operations
(union, intersection, and complementation) transfers from
component automata to temporalized ones: given two classes ${\cal
A}_1$ and ${\cal A}_2$ of automata which are (effectively) closed
under Boolean operations, the class ${\cal A}_1( {\cal A}_2)$ of
temporalized automata is (effectively) closed under Boolean
operations.
\end{theorem}

\begin{proof}

Let $X,Y \in {\cal A}_1( {\cal A}_2)$.

\medskip \noindent {\bf Union} \ \ \ We must provide an automaton
$A \in {\cal A}_1( {\cal A}_2)$ that recognizes the language
${\cal L}(X) \cup {\cal L}(Y)$. Define $A = (X^{\uparrow} \cup
Y^{\uparrow})^\downarrow$. We show that ${\cal L}(A) = {\cal L}(X)
\cup {\cal L}(Y)$. Let $x = (\mathbb{N},<,g) \in {\cal L}(A)$.
Hence, there is  $y = (\mathbb{N},<,V) \in {\cal L}(A^\uparrow) =
{\cal L}(X^\uparrow \cup Y^\uparrow) = {\cal L}(X^\uparrow) \cup
{\cal L}(Y^\uparrow)$ such that, for every $i \in \mathbb{N}$,
$g(i) \in {\cal L}(V(i))$. Suppose $y \in {\cal L}(X^\uparrow)$.
It follows that $x \in {\cal L}(X)$. Hence $x \in {\cal L}(X) \cup
{\cal L}(Y)$. Similarly if $y \in {\cal L}(Y^\uparrow)$.
Conversely, suppose that $x = (\mathbb{N}, <,g) \in {\cal L}(X)
\cup {\cal L}(Y)$. If $x \in {\cal L}(X)$, then there is $y =
(\mathbb{N},<,V) \in {\cal L}(X^\uparrow)$ such that, for every $i
\in \mathbb{N}$, $g(i) \in {\cal L}(V(i))$. Hence, $y \in {\cal
L}(X^\uparrow) \cup {\cal L}(Y^\uparrow) = {\cal L}(X^\uparrow
\cup Y^\uparrow) = {\cal L}(A^\uparrow)$. It follows that $x \in
{\cal L}(A)$. Similarly if $x \in {\cal L}(Y)$.

\medskip \noindent {\bf Complementation} \ \ \
We must provide an automaton $A \in {\cal A}_1( {\cal A}_2)$ that
recognizes the language ${\cal S}({\cal S}(\Sigma)) \setminus
{\cal L}(X)$. Given Lemma~\ref{lm:part}, we may assume that
$\{{\cal L}(Z) \sep Z \in \Gamma(\Sigma)\}$ forms a partition of
${\cal S}(\Sigma)$. We define $A = ({\cal S}(\Gamma(\Sigma))
\setminus X^{\uparrow})^\downarrow$. We show that ${\cal L}(A) =
{\cal S}({\cal S}(\Sigma)) \setminus {\cal L}(X)$. Let $x =
(\mathbb{N},<,g) \in {\cal L}(A)$. Hence, there exists $y =
(\mathbb{N},<,V) \in {\cal L}(A^\uparrow) = {\cal
S}(\Gamma(\Sigma)) \setminus {\cal L}(X^{\uparrow})$ such that,
for every $i \in \mathbb{N}$, $g(i) \in {\cal L}(V(i))$. Suppose,
by contradiction, that $x \in {\cal L}(X)$. It follows that there
exists $z = (\mathbb{N},<,V') \in {\cal L}(X^{\uparrow})$ such
that, for every $i \in \mathbb{N}$, $g(i) \in {\cal L}(V'(i))$.
Hence, for every $i \in \mathbb{N}$, $g(i) \in {\cal L}(V(i)) \cap
{\cal L}(V'(i))$. Since, for every $i \in \mathbb{N}$, ${\cal
L}(V(i)) \cap {\cal L}(V'(i)) = \emptyset$ whenever $V(i) \not =
V'(i)$, we conclude that $V(i) = V'(i)$. Hence $V = V'$ and thus
$y=z$. This is a contradiction since $y$ and $z$ belong to
disjoint sets. It follows that $x \in {\cal S}({\cal S}(\Sigma))
\setminus {\cal L}(X)$.

We now prove the opposite direction. Let $x = (\mathbb{N},<,g) \in
{\cal S}({\cal S}(\Sigma)) \setminus {\cal L}(X)$. It follows
that, for every $y = (\mathbb{N},<,V) \in {\cal L}(X^\uparrow)$,
there exists $i \in \mathbb{N}$ such that $g(i) \not \in {\cal
L}(V(i))$. Suppose, by contradiction, that $x \in {\cal S}({\cal
S}(\Sigma)) \setminus {\cal L}(A)$. It follows that, for every $z
= (\mathbb{N},<,V) \in {\cal L}(A^\uparrow) = {\cal
S}(\Gamma(\Sigma)) \setminus {\cal L}(X^\uparrow)$, there exists
$i \in \mathbb{N}$ such that $g(i) \not \in {\cal L}(V(i))$. We
can conclude that, for every $v = (\mathbb{N},<,V) \in {\cal
S}(\Gamma(\Sigma))$, there exists $i \in \mathbb{N}$ such that
$g(i) \not \in {\cal L}(V(i))$. This is a contradiction: since
$\{{\cal L}(Z) \sep Z \in \Gamma(\Sigma)\}$ forms a partition of
${\cal S}(\Sigma)$, for every $i \in \mathbb{N}$, there is $Y_i
\in \Gamma(\Sigma)$ such that $g(i) \in {\cal L}(Y_i)$. We have
that $(\mathbb{N},<,V')$, with $V'(i) = Y_i$, is an element of
${\cal S}(\Gamma(\Sigma))$ and, for every $i \in \mathbb{N}$,
$g(i) \in {\cal L}(V'(i))$. We conclude that $x \in {\cal L}(A)$.

\medskip \noindent {\bf Intersection} \ \ \ It follows from
closure under union and complementation using De~Morgan's laws.
\end{proof}

It is worth noticing that if $A = (X^{\uparrow} \cap
Y^{\uparrow})^\downarrow$, then ${\cal L}(A) \subseteq {\cal L}(X)
\cap {\cal L}(Y)$, while the opposite inclusion ${\cal L}(X) \cap
{\cal L}(Y) \subseteq {\cal L}(A)$ does not hold in general. We
give a simple counterexample. Let $\Gamma(\Sigma) = \{B,C\}$,
$X^\uparrow$ be the automaton accepting sequences starting with
the symbol $B$, and $Y^\uparrow$ be the automaton accepting
strings starting with the symbol $C$. Then, ${\cal L}(
X^{\uparrow} \cap Y^{\uparrow}) = \emptyset$ and hence ${\cal
L}(A) = \emptyset$. Let $\Sigma = \{a,b\}$, $B$ be the automaton
accepting sequences with an odd number of symbols $a$, and $C$ be
the automaton recognizing sequences with a prime number of symbols
$a$. ${\cal L}(X) \cap {\cal L}(Y)$ contains, for instance, a
combined structure starting with a sequence with exactly 13
occurrences of symbol $a$, and hence it is not empty.

\medskip

We now focus on the problem of establishing whether decidability
transfers from component automata to temporalized ones. Given $A
\in {\cal A}_1( {\cal A}_2)$, it is easy to see that a sufficient
condition for ${\cal L}(A) = \emptyset$ is that ${\cal
L}(A^\uparrow) = \emptyset$. However, this condition is not
necessary, since ${\cal L}(A) = \emptyset$ may depend on the fact
that some ${\cal A}_2$-automata labeling $A$ accept the empty
language. However, if we know that $A$ is labeled with ${\cal
A}_2$-automata recognizing non-empty languages, then the condition
${\cal L}(A^\uparrow) = \emptyset$ is both necessary and
sufficient for ${\cal L}(A) = \emptyset$. In the following
theorem, we take advantage of these considerations to devise an
algorithm that checks emptiness for temporalized automata.

\begin{theorem} \thC{Transfer of decidability} \label{th:tradec}
\noindent Decidability transfers from component automata to
temporalized ones: given two decidable classes  of automata ${\cal
A}_1$ and ${\cal A}_2$, the class ${\cal A}_1( {\cal A}_2)$ of
temporalized automata is decidable.
\end{theorem}

\begin{proof} Let $A$ be a temporalized automaton in ${\cal
A}_1( {\cal A}_2)$.  We describe an algorithm that returns $1$ if
${\cal L}(A) = \emptyset$ and $0$ otherwise.

\begin{description} \item{\bf Step 1} Verify whether ${\cal
L}(A^\uparrow) = \emptyset$ using the algorithm that checks
emptiness for ${\cal A}_1$. If ${\cal L}(A^\uparrow) = \emptyset$,
then return $1$.
\item{\bf Step 2} For every $X \in \Gamma(\Sigma)$, if ${\cal
L}(X) = \emptyset$ (this test can be performed by exploiting the
algorithm that checks emptiness for ${\cal A}_2$), then remove
every transition of the form $(q_1,X,q_2)$ from the transition
relation of $A$.
\item{\bf Step 3} Let $B$ be the temporalized automaton obtained
from $A$ after Step 2. Check, using the emptiness algorithm for
${\cal A}_1$, whether ${\cal L}(B^\uparrow) = \emptyset$. If
${\cal L}(B^\uparrow) = \emptyset$, then return $1$, else return
$0$.
\end{description}

The algorithm always terminates returning either $1$ or $0$. We
prove that the algorithm returns $1$ if and only if ${\cal L}(A) =
\emptyset$. Suppose that the algorithm returns $1$. If ${\cal
L}(A^\uparrow) = \emptyset$, then ${\cal L}(A) = \emptyset$.
Suppose now that ${\cal L}(A^\uparrow) \not = \emptyset$ and
${\cal L}(B^\uparrow) = \emptyset$. Note that ${\cal L}(A) = {\cal
L}(B)$, since $B$ is obtained from $A$ by cutting off automata
accepting the empty language. Assume, by contradiction, that there
is $x \in {\cal L}(A)$. Since ${\cal L}(A) = {\cal L}(B)$, we have
that $x \in {\cal L}(B)$. Hence ${\cal L}(B)$ in not empty. Since
${\cal L}(B^\uparrow) = \emptyset$, we have that ${\cal L}(B)$ is
empty which is a contradiction. Hence ${\cal L}(A) = \emptyset$.
Suppose now that the algorithm returns $0$. Then ${\cal
L}(B^\uparrow)$ contains at least one element, say $x =
(\mathbb{N},<,V)$. Since $B$ uses only non-empty ${\cal
A}_2$-automata as alphabet symbols, we have that, for every $i \in
\mathbb{N}$, ${\cal L}(V(i)) \neq \emptyset$. Hence $y =
(\mathbb{N},<,g)$, with $g$ such that, for every $i \in
\mathbb{N}$, $g(i)$ equals to some element of ${\cal L}(V(i))$, is
an element of ${\cal L}(A)$. Hence ${\cal L}(A) \not = \emptyset$
\end{proof}

Finally, we consider the problem of establishing whether
expressive equivalence with respect to temporal logics transfers
from component automata to temporalized ones. We first state a
partition lemma for temporalized logics. The proof is similar to
the one of Lemma~\ref{lm:part}, and thus omitted.

\begin{lemma} \thC{Partition Lemma for temporalized logics}
\label{lm:part2} \noindent Let $\varphi$ be a temporalized formula
of $\tl{T_1}{T_2}$ and $\alpha_1, \ldots, \alpha_n$ be the maximal
$\logic{T_2}$-formulas of $\varphi$. Then, there exists a finite
set $\Lambda$ of $\logic{T_2}$-formulas such that:
\begin{enumerate}
\item
the set $\{{\cal M}(\alpha) \sep \alpha \in \Lambda\}$ is a
partition of $\bigcup_{i=1}^{n} {\cal M}(\alpha_i)$, and
\item
the formula $\varphi'$ obtained by replacing every
$\logic{T_2}$-formula $\alpha_i$ in $\varphi$ with $\bigvee
\{\alpha \sep \alpha \in \Lambda \mbox{ and } {\cal M}(\alpha)
\cap {\cal M}(\alpha_i) \neq \emptyset\}$ is equivalent to
$\varphi$, i.e., ${\cal M}(\varphi) = {\cal M}(\varphi')$.
\end{enumerate}
\end{lemma}

The following theorem shows that expressive equivalence with
respect to temporal logics transfers from component automata to
temporalized ones.

\begin{theorem} \thC{Transfer of expressive equivalence w.r.t.\
temporal logic} \label{th:traexp} \noindent Expressive equivalence
w.r.t.\ temporal logic transfers from component automata to
temporalized ones: if ${\cal A}_1 \leftrightarrows \logic{T_1}$
${\cal A}_2 \leftrightarrows \logic{T_2}$, and ${\cal A}_2$ is
closed under Boolean operations, then ${\cal A}_1( {\cal A}_2)
\leftrightarrows \tl{T_1}{T_2}$.
\end{theorem}

\begin{proof}
We first prove that ${\cal A}_1( {\cal A}_2) \rightarrow
\tl{T_1}{T_2}$. Let $A \in {\cal A}_1( {\cal A}_2)$ be a
temporalized automaton over $\Gamma(\Sigma) = \{X_1, \ldots ,
X_n\} \subseteq {\cal A}_2$. We have to find a temporalized
formula $\varphi_A \in  \tl{T_1}{T_2}$ such that ${\cal L}(A) =
{\cal M}(\varphi_A)$. Since ${\cal A}_2$ is closed under Boolean
operations, by exploiting Lemma~\ref{lm:part}, we may assume that
$\{{\cal L}(X_1), \ldots , {\cal L}(X_n)\}$ partitions ${\cal
S}(\Sigma)$. Since ${\cal A}_1 \rightarrow \logic{T_1}$, there
exists a translation $\tau_1$ from ${\cal A}_1$-automata to
$\logic{T_1}$-formulas such that, for every $X \in {\cal A}_1$,
${\cal L}(X) = {\cal M}(\tau_1(X))$. Let $\varphi_{A^\uparrow} =
\tau_1(A^\uparrow)$. The formula $\varphi_{A^\uparrow}$ uses
proposition letters in $\{P_{X_1}, \ldots , P_{X_n}\}$. Moreover,
since ${\cal A}_2 \rightarrow \logic{T_2}$, there exists a
translation $\sigma_1$ from ${\cal A}_2$-automata to
$\logic{T_2}$-formulas such that, for every $X \in {\cal A}_2$,
${\cal L}(X) = {\cal M}(\sigma_1(X))$. For every $1 \leq i \leq
n$, let $\varphi_{X_i} = \sigma_1(X_i)$. For every proposition
letter $P_{X_i}$ appearing in $\varphi_{A^\uparrow}$, replace
$P_{X_i}$ by $\varphi_{X_i}$ in $\varphi_{A^\uparrow}$. Let
$\varphi_A$ be the resulting formula. It is immediate to see that
$\varphi_A \in \tl{T_1}{T_2}$. We prove that ${\cal L}(A) = {\cal
M}(\varphi_A)$.

\medskip \noindent
($\subseteq$) \ \ \ Let $x = (\mathbb{N},<,g) \in {\cal L}(A)$.
This implies that there exists $x^\uparrow = (\mathbb{N},<,V) \in
{\cal S}(\Gamma(\Sigma))$ such that $x^\uparrow \in {\cal
L}(A^\uparrow)$ and, for every $i \in \mathbb{N}$, $g(i) \in {\cal
L}(V(i))$.  Since ${\cal L}(A^\uparrow) = {\cal
M}(\varphi_{A^\uparrow})$, we have that $x^\uparrow \in {\cal
M}(\varphi_{A^\uparrow})$.  We prove that, for every $i \in
\mathbb{N}$ and $j \in \{1, \ldots , n\}$, $x^\uparrow, i \models
P_{X_j}$ if and only if $x,i \models \varphi_{X_j}$. Let $i \in
\mathbb{N}$ and $j \in \{1, \ldots , n\}$. We know that
$x^\uparrow, i \models P_{X_j}$ if and only if $V(i) = X_j$. We
first prove that $V(i) = X_j$ if and only if $g(i) \in {\cal
L}(X_j)$. The left to right direction immediately follows since
$g(i) \in {\cal L}(V(i))$. We prove the right to left direction by
contradiction. Suppose $g(i) \in {\cal L}(X_j)$ and $V(i) = X_k
\neq X_j$. Hence $g(i) \in {\cal L}(V(i)) = {\cal L}(X_k)$ and
thus $g(i) \in {\cal L}(X_j) \cap {\cal L}(X_k)$, which is a
contradiction, since ${\cal L}(X_j) \cap {\cal L}(X_k) =
\emptyset$. Hence $V(i)  = X_j$. Finally, we have that $g(i) \in
{\cal L}(X_j)$ if and only if $g(i) \in {\cal M}(\varphi_{X_j})$
if and only if $x,i \models \varphi_{X_j}$. Summing up, we have
that $x^\uparrow \in {\cal M}(\varphi_{A^\uparrow})$ and, for
every $i \in \mathbb{N}$ and $j \in \{1, \ldots , n\}$,
$x^\uparrow, i \models P_{X_j}$ if and only if $x,i \models
\varphi_{X_j}$. It follows that $x \in {\cal M}(\varphi_A)$.

\medskip \noindent
($\supseteq$) \ \ \ Let $x = (\mathbb{N},<,g) \in {\cal
M}(\varphi_A)$. We define $x^\uparrow = (\mathbb{N},<,V) \in {\cal
S}(\Gamma(\Sigma))$ in such a way that, for every $i \in
\mathbb{N}$, $V(i) = X_j$ if and only if $g(i) \in {\cal
M}(\varphi_{X_j}) = {\cal L}(X_j)$. Notice that $V(i)$ is always
and univocally defined, since $\{{\cal L}(X_1), \ldots , {\cal
L}(X_n)\}$ partitions ${\cal S}(\Sigma)$. We prove that, for every
$i \in \mathbb{N}$ and $j \in \{1, \ldots, n\}$, we have that
$x^\uparrow, i \models P_{X_j}$ if and only if $x,i \models
\varphi_{X_j}$. Let $i \in \mathbb{N}$ and $j \in \{1, \ldots ,
n\}$. We know that $x^\uparrow, i \models P_{X_j}$ if and only if
$V(i) = X_j$. We first prove that $V(i) = X_j$ if and only if
$g(i) \in {\cal L}(X_j)$. The left to right direction immediately
follows by definition of $x^\uparrow$.The right to left direction
follows since ${\cal L}(X_j) \cap {\cal L}(X_k) = \emptyset$
whenever $k \neq j$. Finally, $g(i) \in {\cal L}(X_j)$ if and only
if $g(i) \in {\cal M}(\varphi_{X_j})$ if and only if $x,i \models
\varphi_{X_j}$. Summing up, we have that $x^\uparrow \in {\cal
M}(\varphi_{A^\uparrow}) = {\cal L}(A^\uparrow)$ and, for every $i
\in \mathbb{N}$, $g(i) \in {\cal M}(\varphi_{X_j}) = {\cal
M}(\varphi_{V(i)}) = {\cal L}(V(i))$. Therefore, $x \in {\cal
L}(A)$.

\medskip
\noindent We now prove that $\tl{T_1}{T_2} \rightarrow {\cal A}_1(
{\cal A}_2)$. Let $\varphi \in \tl{T_1}{T_2}$ be a temporalized
formula. We have to find a temporalized automaton $A_\varphi \in
{\cal A}_1( {\cal A}_2)$ such that ${\cal M}(\varphi) = {\cal
L}(A_\varphi)$. Let $\alpha_1, \ldots, \alpha_n$ be the maximal
$\logic{T_2}$-formulas of $\varphi$.  By exploiting
Lemma~\ref{lm:part2}, we may assume that there exists a finite set
$\Lambda$ of $\logic{T_2}$-formulas such that the set $\{{\cal
M}(\alpha) \sep \alpha \in \Lambda\}$ forms a partition of
$\bigcup_{i=1}^{n} {\cal M}(\alpha_i)$, and every maximal
$\logic{T_2}$-formula $\alpha_i$ in $\varphi$ has the form
$\bigvee \{\alpha \sep \alpha \in \Lambda \mbox{ and } {\cal
M}(\alpha) \cap {\cal M}(\alpha_i) \neq \emptyset\}$.

Let $\varphi^\uparrow$ be the formula obtained from $\varphi$ by
replacing every $\logic{T_2}$-formula $\alpha \in \Lambda$
appearing in $\varphi$ with proposition letter $P_\alpha$ and by
adding to the resulting formula the conjunct $P_\beta \Or \neg
P_\beta$, where $\beta$ is the $\logic{T_2}$-formula $\neg
\bigvee_{i=1}^{n} \alpha_i$. Let ${\cal Q} = {\{P_{\alpha} \sep
\alpha \in \Lambda \cup \{\beta\}\}}$ be the set of proposition
letters of $\varphi^\uparrow$. Since $\logic{T_1}\rightarrow {\cal
A}_1$, there exists a translation $\tau_2$ from
$\logic{T_1}$-formulas to ${\cal A}_1$-automata such that, for
every $\psi \in \logic{T_1}$, ${\cal M}(\psi) = {\cal
L}(\tau_2(\psi))$. Let $A_{\varphi^\uparrow} =
\tau_2(\varphi^\uparrow)$. The automaton $A_{\varphi^\uparrow}$
labels its transitions with symbols in $2^{\cal Q}$. Moreover,
since $\logic{T_2}\rightarrow {\cal A}_2 $, there exists a
translation $\sigma_2$ from $\logic{T_2}$-formulas to ${\cal
A}_2$-automata such that, for every $\psi \in \logic{T_2}$, ${\cal
M}(\psi) = {\cal L}(\sigma_2(\psi))$. For every $\alpha \in
\Lambda \cup \{\beta\}$, let $A_{\alpha} = \sigma_2(\alpha)$.
Finally, let $A_\varphi$ be the automaton obtained by replacing
every label $X \subseteq {\cal Q}$ on a transition of
$A_{\varphi^\uparrow}$ with the ${\cal A}_2$-automaton
$\bigcap_{P_\alpha \in X} A_{\alpha} =
\sigma_2(\bigwedge_{P_\alpha \in X} \alpha)$. We have that
$A_\varphi \in {\cal A}_1( {\cal A}_2)$ and ${\cal L}(A_\varphi) =
{\cal M}(\varphi)$. The proof is similar to the case ${\cal L}(A)
= {\cal M}(\varphi_A)$. Notice that to prove this direction we did
not use the hypothesis of closure under Boolean operations of
${\cal A}_2$.
\end{proof}

The following corollary shows that, whenever $\logic{T_1}
\rightarrow {\cal A}_1$ and $\logic{T_2} \rightarrow {\cal A}_2$,
the decidability problem for $\logic{T_1}(\logic{T_2})$ can be
reduced to the decidability problems for ${\cal A}_1$ and ${\cal
A}_2$.

\begin{corollary}\label{co:dec}
If $\logic{T_1}  \rightarrow {\cal A}_1$, $\logic{T_2} \rightarrow
{\cal A}_2$, and both ${\cal A}_1$ and ${\cal A}_2$ are decidable,
then $\logic{T_1}(\logic{T_2})$ is decidable.
\end{corollary}

Theorems~\ref{th:traclo},~\ref{th:tradec} and~\ref{th:traexp} hold
for automata that operate on finite sequences as well; moreover,
they can be immediately generalized to automata on finite and
infinite trees (definitions of all these classes of automata can
be found in~\cite{T90}). They remain valid for automata on
temporalized structures that mix sequences and trees.

Corollary~\ref{co:dec} allows one to prove the decidability of
many temporalized logics. For instance, it is well-known that
$\qltl$ (and all its fragments) over infinite sequences can be
embedded into B\"uchi sequence automata, $\qdctls$ (and all its
fragments) over infinite $k$-ary trees can be embedded into Rabin
$k$-ary tree automata, and both B\"uchi sequence and Rabin $k$-ary
tree automata are decidable. Moreover, $\qltl$ (and all its
fragments) over finite sequences can be embedded into finite
sequence automata, $\qdctls$ (and all its fragments) over finite
$k$-ary trees can be embedded into finite $k$-ary tree automata,
and both finite sequence and finite $k$-ary tree automata are
decidable. From Corollary~\ref{co:dec}, it follows that any
temporalized logic $\tl{\logic{T_1}} {\logic{T_2}}$, where
$\logic{T_1}$ and $\logic{T_2}$ are (fragments of) $\qltl$ or
$\qdctls$, interpreted over either finite or infinite sequence or
tree structures, are decidable. As a matter of fact, the
decidability of $\tl{\ptl}{\ptl}$ over infinite sequences of
infinite sequences was already proved in~\cite{FG92} following a
different approach.

\section{Temporalized logics and automata for time granularity}
\label{sec:timegran}

In the following, we use temporalized automata to find the
(combined) temporal logic counterparts of the monadic second-order
theories of downward and upward layered structures. Both results
rest on an alternative view of DULSs and UULSs as infinite
sequences of $k$-ary trees of a suitable form. More precisely,
DULSs can be viewed as infinite sequences of infinite $k$-ary
trees, while UULSs can be interpreted as infinite sequences of
finite {\em increasing} $k$-ary trees. In Section~\ref{sec:DULS}
we provide the monadic second-order theory of DULSs with an
expressively complete and elementarily decidable temporalized
logic counterpart by exploiting a temporalization of B\"uchi and
Rabin automata. Then, in Section~\ref{sec:UULS}, we define a
suitable combination of B\"uchi and finite tree automata and use
it to obtain a combined temporal logic which is both elementarily
decidable and expressively complete with respect to the monadic
second-order theory of UULSs. It is worth remarking that, unlike
the case of DULSs, the combined model we use to encode an UULS
differs from that of temporalization since the innermost submodels
are {\em not} independent from the outermost top-level model.

The monadic second-order language for time granularity ${\rm
MSO}_{\cal P}[<, (\downarrow_i)_{i=0}^{k-1}]$ is defined as
follows.

\begin{definition} \thC{Monadic second-order language}
\noindent Let ${\rm MSO}_{\cal P}[<, (\downarrow_i)_{i=0}^{k-1}]$
be the second-order language with equality  built up as follows:
(i) {\em atomic formulas\/} are of the forms $x = y$, $x < y$,
$\downarrow_i(x) = y$, $x \in X$ and $x \in P$, where $0 \leq i
\leq k-1$, $x$, $y$ are individual variables, $X$ is a set
variable, and $P \in {\cal P}$; (ii) {\em formulas\/} are built up
starting from atomic formulas by means of the Boolean connectives
$\neg$ and $\wedge$, and the quantifier $\exists$ ranging over
both individual and set variables. \dqed
\end{definition}

We interpret ${\rm MSO}_{\cal P}[<, (\downarrow_i)_{i=0}^{k-1}]$
over DULSs and UULSs. For all $i \geq 0$, let $T^i = \{j_i \sep j
\geq 0\}$. A $\cal P$-labeled $k$-refinable DULS is a tuple
$\langle \bigcup_{i \geq 0} T^i, (\downarrow_i)_{i=0}^{k-1}, <,
(P)_{P \in {\cal P}} \rangle$. Part of a $2$-refinable DULS is
depicted in Figure~\ref{fig:down}. A DULS can be viewed as an
infinite sequence of complete $k$-ary infinite trees, each one
rooted at a point of $T^0$. The sets in $\{T^i\}_{i \geq 0}$ are
the layers of the trees, $\downarrow_i$ is a projection function
such that $\downarrow_i(a_b) = c_d$ if and only if $d = b+1$ and
$c = a \cdot k + i$, with $i = 0, \ldots, k-1$, $<$ is a total
ordering over $\bigcup_{i \geq 0} T^i$ given by the {\em preorder}
(root-left-right) visit of the nodes (for elements belonging to
the same tree) and by the total linear ordering of trees (for
elements belonging to different trees), and, for all $P \in {\cal
P}$, $P$ is the set of points in $\bigcup_{i \geq 0} T^i$ labeled
with letter $P$. A $\cal P$-labeled $k$-refinable UULS is a tuple
$\langle \bigcup_{i \geq 0} T^i, (\downarrow_i)_{i=0}^{k-1}, <,
(P)_{P \in {\cal P}} \rangle$.  Part of a $2$-refinable UULS is
depicted in Figure~\ref{fig:up}. An UULS can be viewed as a
$k$-ary infinite tree generated from the leaves. The sets in
$\{T^i\}_{i \geq 0}$ represent the layers of the tree,
$\downarrow_i$ is a projection function such that
$\downarrow_i(a_0) = \bot$, for all $a$, and $\downarrow_i(a_b) =
c_d$ if and only if $b > 0$, $b = d+1$ and $c = a \cdot k + i$,
with $i = 0, \ldots, k-1$, $<$ is the total ordering of
$\bigcup_{i \geq 0} T^i$ given by the {\em inorder}
(left-root-right) visit of the nodes, and, for all $P \in {\cal
P}$, $P$ is the set of points in $\bigcup_{i \geq 0} T^i$ labeled
with letter $P$. Given a formula $\varphi \in {\rm MSO}_{\cal
P}[<, (\downarrow_i)_{i=0}^{k-1}]$, we denote by ${\cal
M}(\varphi)$ the set of models of $\varphi$.

For technical reasons, it is convenient to work with a different,
but equivalent, monadic second-order logic over DULSs that
replaces the total ordering $<$ by two partial orderings $<_1$ and
$<_2$ defined as follows. Let $t$ be a DULS. According to the
interpretation of DULSs as tree sequences, we define $x <_1 y$ if
and only if $x$ is the root of some tree $t_i$ of $t$, $y$ is the
root of some tree $t_j$ of $t$, and $i<j$ over natural numbers.
Moreover, $x <_2 y$ if and only if $y$ is different from $x$ and
$y$ belongs to the tree rooted at $x$. In a similar way, it is
convenient to work with a different, but equivalent, monadic
second-order logic over UULSs that replaces the total ordering $<$
with a partial ordering $<_{pre}$ such that $x <_{pre} y$ if and
only if $y$ is different from $x$ and $y$ belongs to the tree
rooted at $x$.

\subsection{Downward unbounded layered structures}
\label{sec:DULS}

\begin{figure}[t]
\begin{center}
\begin{picture}(360,90)
\put(-10,-10){\begin{picture}(170,90) \put(90,90){\circle*{3}}
\put(90,90){\line(1,0){250}} 
\multiput(50,70)(80,0){2}{\circle*{3}}
\put(50,70){\line(2,1){40}} \put(130,70){\line(-2,1){40}}
\multiput(30,50)(40,0){4}{\circle*{3}}
\multiput(30,50)(80,0){2}{\line(1,1){20}}
\multiput(70,50)(80,0){2}{\line(-1,1){20}}
\multiput(20,30)(20,0){8}{\circle*{3}}
\multiput(20,30)(40,0){4}{\line(1,2){10}}
\multiput(40,30)(40,0){4}{\line(-1,2){10}}
\multiput(15,20)(20,0){8}{\line(1,2){5}}
\multiput(25,20)(20,0){8}{\line(-1,2){5}}
\multiput(14,10)(20,0){8}{$\ldots$} \put(90,100){$t_0$}
\put(260,100){$t_1$}
\end{picture}}
\put(160,-10){\begin{picture}(170,90) \put(90,90){\circle*{3}}
\multiput(50,70)(80,0){2}{\circle*{3}} \put(50,70){\line(2,1){40}}
\put(130,70){\line(-2,1){40}}
\multiput(30,50)(40,0){4}{\circle*{3}} \put(175,30){$...$}
\put(175,50){$...$} \put(175,70){$...$} \put(175,90){$...$}
\multiput(30,50)(80,0){2}{\line(1,1){20}}
\multiput(70,50)(80,0){2}{\line(-1,1){20}}
\multiput(20,30)(20,0){8}{\circle*{3}}
\multiput(20,30)(40,0){4}{\line(1,2){10}}
\multiput(40,30)(40,0){4}{\line(-1,2){10}}
\multiput(15,20)(20,0){8}{\line(1,2){5}}
\multiput(25,20)(20,0){8}{\line(-1,2){5}}
\multiput(14,10)(20,0){8}{$\ldots$} \put(175,10){$...$}
\end{picture}}
\end{picture}
\caption{\label{fig:treeseq1} A tree sequence.}
\end{center}
\end{figure}
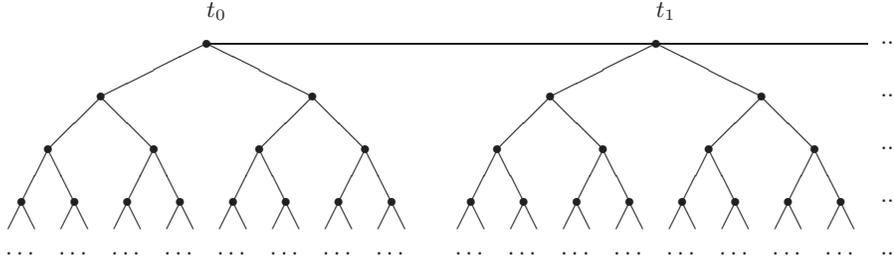

We start with a formalization of the alternative characterization
of DULSs as suitable tree sequences given above. Let ${\cal
T}_k({\cal P})$ be the set of ${\cal P}$-labeled infinite $k$-ary
trees. Let ${\cal S}({\cal T}_k({\cal P}))$ be the set of infinite
sequences of ${\cal P}$-labeled infinite $k$-ary trees, that is,
temporalized models $(\mathbb{N},<,g)$ where $g \, : \, \mathbb{N}
\Imp {\cal T}_k({\cal P})$. ${\cal P}$-labeled DULSs correspond to
tree sequences in ${\cal S}({\cal T}_k({\cal P}))$, and vice
versa. On the one hand, ${\cal P}$-labeled DULS $t$ can be viewed
as an infinite sequence of ${\cal P}$-labeled infinite $k$-ary
trees, whose $i$-th tree, denoted by $t_i$, is the ${\cal
P}$-labeled tree rooted at the $i$-th point $i_0$ of the coarsest
domain $T^0$ of $t$ (cf.\ Figure~\ref{fig:treeseq1}). Such a
sequence can be represented as the temporalized model
$(\mathbb{N},<,g) \in {\cal S}({\cal T}_k({\cal P}))$ such that,
for every $i \in \mathbb{N}$, $g(i) = t_i$. On the other hand, it
is immediate to reinterpret infinite sequences of ${\cal
P}$-labeled infinite $k$-ary trees in terms of ${\cal P}$-labeled
DULSs.

Such a correspondence between DULSs and temporalized models
enables us to use temporalized logics
$\tl{\logic{T_1}}{\logic{T_2}}$, where $\logic{T_1}$ is a linear
time logic and $\logic{T_2}$ is a branching time logic, to express
properties of DULSs. Furthermore, taking advantage of the
correspondence between temporalized logic and automata, we can
equivalently use temporalized automata $\tl{{\cal A}_1}{{\cal
A}_2}$ over DULSs, where ${\cal A}_1$ is a class of sequence
automata and ${\cal A}_2$ is a class of tree automata. In the
following, we will focus on the class ${\cal B}({\cal R}_k)$ of
temporalized automata embedding Rabin $k$-ary tree automata into
B\"uchi sequence automata. We call automata in this class {\em
infinite tree sequence automata}. Since both ${\cal B}$ and ${\cal
R}_k$ are effectively closed under Boolean operations and
decidable, Theorems~\ref{th:traclo} and~\ref{th:tradec} allow us
to conclude that the class ${\cal B}({\cal R}_k)$ of infinite tree
sequence automata is effectively closed under Boolean operations
and decidable as well. The complexity of the emptiness problem for
infinite tree sequence automata is given by the following theorem.

\begin{theorem}
\thC{Complexity of infinite tree sequence automata}
\label{th:compTSA} \noindent The emptiness problem for infinite
tree sequence automata is decidable in polynomial time in the
number of states, and exponential time in the number of accepting
pairs.
\end{theorem}

\begin{proof}
For any given $A \in {\cal B}({\cal R}_k)$, let $n$ be the number
of states of $A$ and $N$ (resp.\ $M$) be the maximum number of
states (resp.\ accepting pairs) of a Rabin tree automaton labeling
transitions in $A$. The emptiness of B\"uchi sequence automata can
be checked in polynomial time in the number of states, while the
emptiness of Rabin tree automata can be verified in polynomial
time in the number of states, and exponential time in the number
of accepting pairs. By applying the algorithm used to test the
emptiness of temporalized automata in the proof of
Theorem~\ref{th:tradec}, we have that the complexity of checking
whether $A$ accepts the empty language is polynomial in $n$ and
$N$, and exponential in $M$.
\end{proof}

The following theorem relates infinite tree sequence automata to
the monadic second-order theory of DULSs.

\begin{theorem}
\thC{Expressiveness of infinite tree sequence automata}
\label{th:expTSA} \noindent Infinite tree sequence automata are as
expressive as the monadic second-order theory of DULSs.
\end{theorem}

\begin{proof}

The proof can be accomplished following a proof strategy that
closely resembles those adopted to prove classical results in the
field, such as, for instance, the proof of B\"uchi's Theorem
(cf.~\cite{T90}). We split it in two parts:

\begin{description}
\item{(a)}
we first show that, for every automaton $A \in {\cal B}({\cal
R}_k)$ over $\Gamma(\Sigma)$, there exists a formula $\varphi_A
\in {\rm MSO}_{{\cal P}_\Sigma}[<_1,<_2,
(\downarrow_i)_{i=0}^{k-1}]$ over ${\cal P}_\Sigma = \{P_a \sep a
\in \Sigma\}$ such that ${\cal L}(A) = {\cal M}(\varphi_A)$;
\item{(b)}
then, we show that, for every formula $\varphi \in {\rm MSO}_{\cal
P}[<_1,<_2, (\downarrow_i)_{i=0}^{k-1}]$ over ${\cal P}$, there
exists an automaton $A_\varphi \in {\cal B}({\cal R}_k)$ over some
$\Gamma(2^{\cal P})$ such that ${\cal M}(\varphi) = {\cal
L}(A_\varphi)$.
\end{description}

We first introduce some auxiliary predicates that can be easily
defined in the monadic second-order logic over DULSs. Let $+ 1$ be
a binary predicate such that $+1(x,y)$ if and only if $x$ and $y$
belong to the coarsest domain and $y$ is the immediate successor
of $x$. We will write $x+1 \in X$ for $\E y (+1(x,y) \And y \in
X)$. Moreover, let ${\tt T}^0(x)$ be a shorthand for ``x belongs
to the coarsest domain'', $0_0 \in X$ be a shorthand for ``the
first element of the coarsest domain belongs to $X$'', and ${\tt
Path}(X,x)$ be a shorthand for the formula stating that ``$X$ is a
path rooted at $x$''.

Let us prove part (a) for $k=2$. The generalization to $k > 2$ is
straightforward. Let $A = (Q,q_0,\Delta,F)$ be a ${\cal B}({\cal
R}_2)$-automaton over $\Gamma(\Sigma)$ (finite subset of ${\cal
R}_2$) accepting tree sequences in ${\cal S}({\cal T}_2(\Sigma))$.
We produce a sentence $\varphi_A \in {\rm MSO}_{{\cal
P}_\Sigma}[<_1,<_2, \proj{0},\proj{1}]$, that involves monadic
predicates in ${\cal P}_\Sigma = \{P_a \sep a \in \Sigma\}$ and is
interpreted over ${\cal S}({\cal T}_2(\Sigma))$, such that ${\cal
L}(A) = {\cal M}(\varphi_A)$. We assume $Q = \{0, \ldots m\}$ and
$q_0 = 0$. For every $Z \in \Gamma(\Sigma)$, let $Z =
(Q_Z,q_{Z}^{0},\Delta_Z,\Gamma_Z)$ over $\Sigma$, with $Q_Z = \{0,
\ldots m_Z\}$, $q_{Z}^{0} = 0$, and $\Gamma_Z =
\{(L_{i}^{Z},U_{i}^{Z}) \sep 1 \leq i \leq r_Z\}$.

The ${\rm MSO}_{{\cal P}_\Sigma}[<_1,<_2, \proj{0},
\proj{1}]$-sentence $\varphi_A$  that corresponds to the automaton
$A$ basically encodes the combined acceptance condition for ${\cal
B}({\cal R}_2)$-automata. The outermost part of the sentence
expresses the existence of an accepting run over the coarsest
layer of the tree sequence for the B\"uchi sequence automaton
$A^\uparrow$. For all $i \in Q$, the second-order variable $X_i$
denotes the set of positions of the run which are associated with
the state $i$, while, for all $Z \in \Gamma(\Sigma)$ the monadic
predicate $Q_Z$ denotes the set of positions of the run that are
labeled with the Rabin tree automaton $Z$. The innermost part
${\tt RAC}(x,Z)$ captures the existence of an accepting run over
the tree rooted at $x$ for the Rabin tree automaton $Z$. For $i
\in Q_Z$, the second-order variable $Y_i$ denotes the set of
positions of the run that are associated with state $i$. The
sentence $\varphi_A$ is defined as follows:

\medskip
\noindent $\begin{array}{l} (\E Q_Z)_{Z \in \Gamma(\Sigma)} (\E
X_i)_{i=0}^{m}  (\bigwedge_{i=0}^{m} \A x (x \in X_i \Imp {\tt
T}^0(x)) \And \\ \bigwedge_{Z \in \Gamma(\Sigma)} \A x (x \in Q_Z
\Imp {\tt T}^0(x)) \And 0_0 \in X_0 \And \bigwedge_{i \neq j} \neg
\E y (y \in X_i \And y \in X_j) \And \\ \A x ({\tt T}^0(x) \Imp
\bigvee_{(i,Z,j) \in \Delta}(x \in X_i \And x \in Q_Z \And x+1 \in
X_j)) \And \\ \bigvee_{i \in F} \A x ({\tt T}^0(x) \Imp \E y ({\tt
T}^0(y) \And x <_1 y \And y \in X_i)) \And \\ \bigwedge_{Z \in
\Gamma(\Sigma)} \A x (x \in Q_Z \Imp {\tt RAC}(x,Z)),
\end{array}$

\medskip
\noindent where ${\tt RAC}(x,Z)$ stands for:

\medskip
\noindent $\begin{array}{l} (\E Y_i)_{i=0}^{m_Z} (\bigwedge_{i=
0}^{m_Z} \A y (y \in Y_i \Imp x \leq_2 y) \And x \in Y_0 \And
\bigwedge_{i \neq j} \neg \E y (y \in Y_i \And y \in Y_j) \And \\
\A y (x \leq_2 y \Imp \bigvee_{(i,a,j_0,j_1) \in \Delta_Z}(y \in
Y_i \And y \in P_a \And  \proj{0}(y) \in Y_{j_0} \And \proj{1}(y)
\in Y_{j_1})) \And \\ \A W ({\tt Path}(W,x) \Imp
\bigvee_{i=0}^{r_Z} (\bigwedge_{j \in L_{i}^{Z}} \E u (u \in W
\And \A v (v \in W \And u <_2 v \Imp v \not \in Y_j)) \And \\
\bigvee_{j \in U_{i}^{Z}} \A u (u \in W \Imp \E v (v \in W \And u
<_2 v \And v \in Y_j))))).
\end{array}$

\medskip

We now prove part (b). Let ${\cal P} = \{P_1, \ldots P_n\}$. To
simplify things, we prove our result for the theory ${\rm
MSO}_{{\cal P}}[<_1,<_2, (\downarrow_i)_{i=0}^{k-1},+1]$ which can
be easily shown to be equivalent to ${\rm MSO}_{{\cal P}}[<_1,<_2,
(\downarrow_i)_{i=0}^{k-1}]$. Given a formula $\varphi \in {\rm
MSO}_{{\cal P}}[<_1,<_2, (\downarrow_i)_{i=0}^{k-1},+1]$, that
involves monadic predicates in ${\cal P}$ and is interpreted over
${\cal P}$-labeled tree sequences in ${\cal S}({\cal T}_k({\cal
P}))$, we build an automaton $A_\varphi \in {\cal B}({\cal R}_k)$
over some $\Gamma(2^{\cal P})$ and accepting in ${\cal S}({\cal
T}_k({\cal P}))$ such that ${\cal L}(A_\varphi) = {\cal
M}(\varphi)$.

As a first step, we show that the ordering relations $<_1$ and
$<_2$ can actually be removed without reducing the expressiveness.
We replace $x <_1 y$ by $${\tt T}^{0}(x) \And {\tt T}^{0}(y) \And
\A X (x + 1 \in X \And \A z (z \in X \Imp z + 1 \in X) \Imp y \in
X)),$$ and $x <_2 y$ by $$\A X ( \bigwedge_{i =0}^{k-1}
\proj{i}(x) \in X \And \A z(z \in X \Imp \bigwedge_{i=0}^{k-1}
\proj{i}(z) \in X) \Imp y \in X).$$ Hence, ${\rm MSO}_{\cal
P}[<_1, <_2, (\proj{i})_{i= 0}^{k-1}, +1]$ is as expressive as
${\rm MSO}_{\cal P}[(\proj{i})_{i=0}^{k-1}, +1]$. Next, we
introduce an expressively equivalent variant of ${\rm MSO}_{\cal
P}[(\proj{i})_{i=0}^{k-1}, +1]$, denoted by ${\rm
MSO}[(\proj{i})_{i = 0}^{k-1}, +1]$, which uses free set variables
$X_i$ in place of predicate symbols $P_i$ and is interpreted over
$\{0,1\}^n$-labeled tree sequences in ${\cal S}({\cal
T}_k(\{0,1\}^n))$. The idea is to encode a set $X \subseteq {\cal
P}$ with the string $i_1 \ldots i_n \in \{0,1\}^n$ such that, for
$j = 1,\ldots,n$,  $i_j = 1$ if and only if $P_j \in X$. We now
reduce ${\rm MSO}[(\proj{i})_{i=0}^{k-1}, +1]$ to a simpler
formalism ${\rm MSO}_{0}[(\proj{i})_{i=0}^{k-1}, +1]$, where {\em
only} second-order variables $X_i$ occur and atomic formulas are
of the forms $X_i \subseteq X_j$  ($X_i$ is a subset of $X_j$),
${\tt Proj}_m(X_i,X_j)$, with $m = 0, \ldots, k-1$ ($X_i$ and
$X_j$ are the singletons $\{x\}$ and $\{y\}$, respectively, and
$\proj{m}(x) = y$), and ${\tt Succ}(X_i,X_j)$  ($X_i$ and $X_j$
are the singletons $\{x\}$ and $\{y\}$, respectively, and $x + 1 =
y$). This step is performed as in the proof of B\"uchi's Theorem.
\begin{figure}[t]
\epsfxsize=1.5in \centerline{\epsfbox {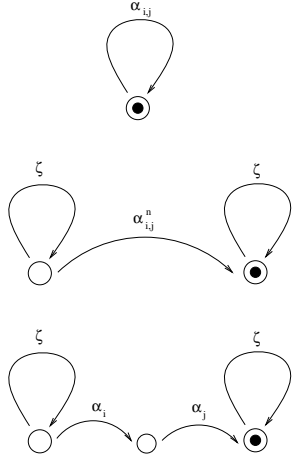}}
\caption{Temporalized automata for atomic formulas.}
\label{fig:atomic}
\end{figure}
Finally, given a ${\rm MSO}_{0}[(\proj{i})_{i=0}^{k-1},
+1]$-formula $\varphi(X_1, \ldots , X_n)$, we prove, by induction
on the structural complexity of $\varphi$, that there exists a
temporalized automaton $A_\varphi$ accepting in ${\cal S}({\cal
T}_k(\{0,1\}^n))$ such that $\mc{M}(\varphi) = \mc{L}(A_\varphi)$.
A corresponding automaton accepting in ${\cal S}({\cal T}_k({\cal
P}))$ can be obtained in the obvious way. As for atomic formulas,
let $\alpha_{i,j}$ be the Rabin tree automaton over $\{0,1\}^n$
for $X_i \subseteq X_j$. The temporalized automaton for $X_i
\subseteq X_j$ is depicted in Figure~\ref{fig:atomic} (top).
Moreover, let $\zeta$ be the Rabin tree automaton over $\{0,1\}^n$
that accepts the singleton set containing a tree labeled with
$0^n$ everywhere, and let $\alpha_{i,j}^{m}$ be the Rabin tree
automaton over $\{0,1\}^n$ for ${\tt Proj}_m(X_i,X_j)$. The
temporalized automaton for ${\tt Proj}_m(X_i,X_j)$ is depicted in
Figure~\ref{fig:atomic} (middle). Finally, let $\alpha_i$ be the
Rabin tree automaton over $\{0,1\}^n$ that accepts the singleton
set containing a tree labeled with $0^{i-1}10^{n-i}$ at the root,
and labeled with $0^n$ elsewhere. The combined automaton for ${\tt
Succ}(X_i,X_j)$ is depicted in Figure~\ref{fig:atomic} (bottom).
The induction step immediately follows from the closure of ${\cal
B}({\cal R}_k)$ automata under Boolean operations and projection.
Closure under Boolean operations has been already shown; closure
under projection can be argued as follows: given a ${\cal B}({\cal
R}_k)$-automaton $A$, the corresponding projected ${\cal B}({\cal
R}_k)$-automaton is obtained by simply projecting every Rabin
automaton that labels some transition of $A$.
\end{proof}

We can exploit infinite tree sequence automata to provide the
(full) second-order theory of DULSs with an expressively complete
and elementarily decidable temporal logic counterpart. First of
all, it is well-known that ${\cal B} \leftrightarrows \qltl$ and
${\cal B} \leftrightarrows \eqltl$, as well as ${\cal R}_k
\leftrightarrows \qdctls$ and ${\cal R}_k \leftrightarrows
\eqdctls$~\cite{E90}. Since Rabin tree automata are closed under
Boolean operations, Theorem~\ref{th:traexp} allows us to conclude
that both $\tl{\qltl}{\qdctls} \leftrightarrows {\cal B}({\cal
R}_k)$ and $\tl{\eqltl}{\eqdctls} \leftrightarrows {\cal B}({\cal
R}_k)$\footnote{It is worth pointing out that the application of
the partition step of Theorem~\ref{th:traexp} to temporal formulas
in $\eqdctls$ generates formulas of the form $\neg \E Q_1 \ldots
\E Q_n \varphi$, where $\varphi$ is a $\dctls$-formula, which do
not belong to the language of $\eqdctls$, because such a language
is not closed under negation. Nevertheless, formulas of the form
$\neg \E Q_1 \ldots \E Q_n \varphi$ can be embedded into Rabin
tree automata as well. The Rabin tree automaton for $\neg \E Q_1
\ldots \E Q_n \varphi$ can indeed be obtained by taking the
complementation of the projection, with respect to $Q_1, \ldots
Q_n$, of the Rabin tree automaton for $\varphi$.}. By applying
Theorem~\ref{th:expTSA}, we have that both $\tl{\qltl}{\qdctls}
\leftrightarrows {\rm MSO}_{\cal P}[<_1,<_2,
(\downarrow_i)_{i=0}^{k-1}]$ and $\tl{\eqltl}{\eqdctls}
\leftrightarrows {\rm MSO}_{\cal P}[<_1,<_2,
(\downarrow_i)_{i=0}^{k-1}]$. Such a result is summarized by the
following theorem.

\begin{theorem}
\thC{Expressiveness of $\tl{\qltl}{\qdctls}$ and
$\tl{\eqltl}{\eqdctls}$} \label{th:expQTL} \noindent
$\tl{\qltl}{\qdctls}$ and $\tl{\eqltl}{\eqdctls}$ are as
expressive as ${\rm MSO}_{\cal P}[<_1,<_2,
(\downarrow_i)_{i=0}^{k-1}]$, when interpreted over DULSs.
\end{theorem}

Furthermore, since ${\rm MSO}_{\cal P}[<_1,<_2,
(\downarrow_i)_{i=0}^{k-1}]$ is decidable, both
$\tl{\qltl}{\qdctls}$ and $\tl{\eqltl}{\eqdctls}$ are decidable.
The next theorem shows that $\tl{\eqltl}{\eqdctls}$ is {\em
elementarily} decidable.

\begin{theorem}
\thC{Complexity of $\tl{\eqltl}{\eqdctls}$} \label{th:compQTL}
\noindent The satisfiability problem for $\tl{\eqltl}{\eqdctls}$
over DULSs is in ELEMENTARY.
\end{theorem}

\begin{proof}
$\tl{\eqltl}{\eqdctls}$ can be decided by embedding it into ${\cal
B}({\cal R}_k)$ automata (such an embedding can be accomplished
following the approach outlined in the proof of
Theorem~\ref{th:traexp}). $\eqltl$ can be elementarily embedded
into B\"uchi sequence automata. Indeed, given an $\eqltl$-formula
$\E Q_1 \ldots \E Q_n \varphi$, the $\ptl$-formula $\varphi$ can
be converted into a B\"uchi sequence automaton $A_\varphi$ of size
$O(2^{|\varphi|})$. A B\"uchi sequence automaton for $\E Q_1
\ldots \E Q_n \varphi$ can be obtained by taking the projection of
$A_\varphi$ with respect to letters $Q_1, \ldots, Q_n$, that is,
by deleting letters $Q_1, \ldots, Q_n$ from the transitions of
$A_\varphi$. The size of the resulting automaton is
$O(2^{|\varphi|})$. Similarly, $\eqdctls$ formulas can be embedded
into Rabin tree automata with a doubly exponential number of
states and a singly exponential number of accepting pairs in the
length of the formula. In particular, as already pointed out, a
Rabin tree automaton for formulas of the form $\neg \E Q_1 \ldots
\E Q_n \varphi$, which are generated by applying the partition
step of Theorem~\ref{th:traexp} to $\eqdctls$ formulas, can be
obtained by taking the complementation of the projection, with
respect to $Q_1, \ldots Q_n$, of the Rabin tree automaton for
$\varphi$. The resulting automaton has elementary size. Hence, any
$\tl{\eqltl}{\eqdctls}$ formula can be converted into an
equivalent ${\cal B}({\cal R}_k)$ automaton of elementary size.
Since ${\cal B}({\cal R}_k)$ automata are elementarily decidable,
we have the thesis.
\end{proof}

We conclude the section by giving some examples of meaningful
timing properties that can be expressed in (fragments of)
$\tl{\eqltl}{\eqdctls}$ interpreted over DULSs. As a first
example, consider the property `$P$ densely holds at some node
$x$' meaning that there exists a path rooted at $x$ such that $P$
holds at each node of the path (notice that such a property
implies that, for every $i \geq 0$, there exists $y \in
\downarrow^i(x)$ such that $P$ holds at $y$, where, for $i \geq
0$, $\downarrow^i(x)$ is the $i$-th layer of the tree rooted at
$x$, but not vice versa). This property can be expressed in
$\tl{\ptl}{\dctls}$ by the formula: $$\Diamond \op{E F E G}P.$$ As
another example, the property `$P$ holds at the origin of every
layer' (or, equivalently, `$P$ holds along the leftmost path of
the first tree of the sequence') can be expressed in
$\tl{\ptl}{\dctls}$ as follows: $$\op{E}(P \And \op{GX_0}P).$$ As
a third example, the property `$P$ holds everywhere on every even
tree' can be encoded in $\tl{\eqltl}{\dctls}$ as follows: $$\E Q(Q
\And \Next \neg Q \And \Box(Q \IImp \Next \Next Q) \And \Box(Q
\Imp \op{AG} P)).$$ Notice that such a property cannot be
expressed in $\tl{\ptl}{\dctls}$, since, as it is well-known,
$\ptl$ cannot express the property `$P$ holds on every even
point'~\cite{W83}. As a last example, the property `$P$ holds
everywhere on every even layer' can be encoded in
$\tl{\ptl}{\eqdctls}$ as follows: $$\Box \E Q(Q \And \op{AX} \neg
Q \And \op{A G}(Q \IImp \op{AXAX} Q) \And \op{AG}(Q \Imp P)).$$
Notice that also this property cannot be expressed in
$\tl{\ptl}{\dctls}$.

Unfortunately, things are not always that easy. As an example, the
property `$P$ holds at exactly one node' can be easily encoded in
(the first-order fragment of) ${\rm MSO}_{\cal P}[<_1,$ $<_2,
(\downarrow_i)_{i=0}^{k-1}]$ by the formula: $\E x (x \in P \And
\A y(y \neq x \Imp y \not \in P))$, while it is not easy at all to
express it in $\tl{\eqltl}{\eqdctls}$. Moreover, since ${\rm
MSO}_{\cal P}[<_1,<_2, (\downarrow_i)_{i=0}^{k-1}]$ is
nonelementarily decidable, while $\tl{\eqltl}{\eqdctls}$ is
elementarily decidable, the translation $\tau$ of ${\rm MSO}_{\cal
P}[<_1,$ $<_2, (\downarrow_i)_{i=0}^{k-1}]$ formulas into
$\tl{\eqltl}{\eqdctls}$ formulas is nonelementary. This means
that, for every $n \in \mathbb{N}$, there exists an ${\rm
MSO}_{\cal P}[<_1,<_2, (\downarrow_i)_{i=0}^{k-1}]$-formula
$\varphi$ such that the length of $\tau(\varphi)$ is greater than
$\kappa(n, |\varphi|)$ (an exponential tower of height $n$).

\subsection{Upward unbounded layered structures}
\label{sec:UULS}

\begin{figure}[t]
\begin{center}
{\begin{picture}(250,90) \multiput(0,50)(80,0){4}{\circle*{3}}
\multiput(80,50)(80,0){3}{\line(0,-1){20}}
\multiput(80,30)(80,0){3}{\circle*{3}}
\multiput(160,30)(80,0){2}{\line(-1,-1){20}}
\multiput(160,30)(80,0){2}{\line(1,-1){20}}
\multiput(140,10)(40,0){4}{\circle*{3}}
\multiput(220,10)(40,0){2}{\line(-1,-1){10}}
\multiput(220,10)(40,0){2}{\line(1,-1){10}}
\multiput(210,0)(20,0){4}{\circle*{3}} \put(0,50){\line(1,0){240}}
\put(260,50){$\ldots$} \put(0,60){$0_0$} \put(80,60){$0_1$}
\put(160,60){$0_2$} \put(240,60){$0_3$} \put(85,30){$1_0$}
\put(165,30){$1_1$} \put(245,30){$1_2$} \put(148,10){$2_0$}
\put(188,10){$3_0$} \put(228,10){$2_1$} \put(268,10){$3_1$}
\put(208,-15){$4_0$} \put(228,-15){$5_0$} \put(248,-15){$6_0$}
\put(268,-15){$7_0$}
\end{picture}}
\vspace{1cm} \caption{\label{fig:treeseq3} Mapping an UULS into an
increasing tree sequence.}
\end{center}
\end{figure}
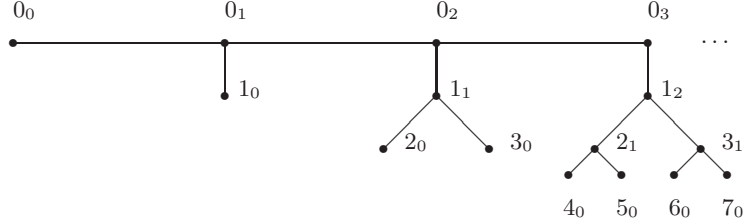

We start by giving an alternative characterization of UULSs in
terms of tree sequences. To this end, we need to introduce the
notions of almost $k$-ary tree and of increasing tree sequence. An
{\em almost $k$-ary finite tree} is a complete finite tree whose
root has exactly $k-1$ sons $0, \ldots, k-2$, each of them is the
root of a complete finite $k$-ary tree. Let ${\cal H}_k({\cal P})$
be the set of $\cal P$-labeled almost $k$-ary finite trees. A
${\cal P}$-labeled {\em increasing $k$-ary tree sequence} (ITS,
for short) is a tree sequence such that, for every $i \in
\mathbb{N}$, the $i$-th tree of the sequence is a ${\cal
P}$-labeled almost $k$-ary tree of height $i$ (cf.\
Figure~\ref{fig:treeseq3}). A ${\cal P}$-labeled ITS can be
represented as a temporalized model $(\mathbb{N},<,g)$ such that,
for every $i \in \mathbb{N}$, $g(i)$ is the $i$-th tree of the
sequence. Let $ITS_k({\cal P})$ be the set of ${\cal P}$-labeled
$k$-ary ITSs. It is worth noting that $ITS_k({\cal P})$ is {\em
not} the class ${\cal H}_k({\cal P}$ of temporalized models
embedding almost $k$-ary finite trees into infinite sequences: an
increasing tree sequence is a particular sequence of almost
$k$-ary finite trees, but a sequence of almost $k$-ary finite
trees is not necessary increasing, and thus $ITS_k({\cal P})
\subsetneq {\cal S}({\cal H}_k({\cal P}))$.

It is not difficult to show that a ${\cal P}$-labeled UULS
corresponds to a ${\cal P}$-labeled ITS, and vice versa. As
already pointed out, an UULS can be viewed as an infinite complete
$k$-ary tree generated from the leaves. The corresponding tree
sequence can be obtained starting from the first point of the
finest layer of the UULS and climbing up along the leftmost path
of the structure. The $i$-th tree in the sequence is obtained by
taking the tree rooted at the $i$-th point of the leftmost path,
and by deleting from it the subtree rooted at the leftmost son of
its root. More precisely, let $t$ be a $k$-ary UULS. For every
node $x$ in $t$, we define $t_x$ to be the finite complete $k$-ary
tree rooted at $x$. For every $i \geq 0$, let $\hat{t}_{0_i}$ be
the almost $k$-ary finite tree obtained from $t_{0_i}$ by
deleting, whenever $i>0$, the subtree $t_{0_{i-1}}$ from it. The
ITS $(\mathbb{N},<,g)$ associated with the UULS $t$ is obtained by
defining, for every $i \geq 0$, $g(i) = \hat{t}_{0_i}$. The
embedding of a binary UULS into a binary ITS is depicted in
Figure~\ref{fig:treeseq3}. Similarly, ITSs can be reinterpreted in
terms of UULSs.

On the basis of such a correspondence between UULSs and ITSs, we
can use temporalized logics $\tl{\logic{T_1}}{\logic{T_2}}$, where
$\logic{T_1}$ is a linear time logic and $\logic{T_2}$ is a
branching time logic, to express properties of UULSs. More
precisely, we interpret $\tl{\logic{T_1}}{\logic{T_2}}$ over
${\cal S}({\cal H}_k({\cal P}))$, but, since we are interested in
increasing tree sequences, we study the logical properties of
$\tl{\logic{T_1}}{\logic{T_2}}$, such as expressiveness and
decidability, with respect to the proper subset $ITS_k({\cal P})$.
Temporalized automata $\tl{{\cal A}_1}{{\cal A}_2}$ over UULSs can
be defined in a similar way. Once again, we consider automata in
$\tl{{\cal A}_1}{{\cal A}_2}$ accepting in ${\cal S}({\cal
H}_k(\Sigma))$, but, since we are interested in increasing tree
sequences, we study the relevant properties of $\tl{{\cal
A}_1}{{\cal A}_2}$, such as closure under Boolean operations,
expressiveness, and decidability, with respect to the proper
subset $ITS_k(\Sigma)$. In the following, we will focus on the
class ${\cal B}({\cal C}_k)$ of temporalized automata embedding
almost $k$-ary finite tree automata into B\"uchi sequence
automata. We call automata in ${\cal B}({\cal C}_k)$ {\em finite
tree sequence automata}.

Since both ${\cal B}$ and ${\cal C}_k$ are effectively closed
under Boolean operations and decidable, Theorems~\ref{th:traclo}
and~\ref{th:tradec} allows us to conclude that ${\cal B}({\cal
C}_k)$ is effectively closed under Boolean operations and
decidable. We show that ${\cal B}({\cal C}_k)$-automata are closed
under Boolean operations over the set $ITS_k(\Sigma)$ as well. Let
$A,B \in {\cal B}({\cal C}_k)$. We show that:

\begin{itemize}
\item
there exists $C \in {\cal B}({\cal C}_k)$ such that $${\cal L}(C)
\cap ITS_k(\Sigma) = ITS_k(\Sigma) \setminus {\cal L}(A) \ \
(complementation);$$
\item
there exists $C \in {\cal B}({\cal C}_k)$ such that $${\cal L}(C)
\cap ITS_k(\Sigma) = ({\cal L}(A) \cup {\cal L}(B)) \cap
ITS_k(\Sigma) \ \ (union);$$
\item
there exists $C \in {\cal B}({\cal C}_k)$ such that $${\cal L}(C)
\cap ITS_k(\Sigma) = ({\cal L}(A) \cap {\cal L}(B)) \cap
ITS_k(\Sigma) \ \ (intersection).$$
\end{itemize}

As it can be easily checked, it suffices to set $C = \overline{A}$
in case of complementation, $C = A \cup B$ in the case of union,
and $C = A \cap B$ in the case of intersection.

The following theorem relates finite tree sequence automata to the
monadic second-order theory of UULSs.

\begin{theorem}
\label{th:expkFTSA} \thC{Expressiveness of finite tree sequence
automata} \noindent Finite tree sequence automata are as
expressive as the monadic second-order theory of UULSs.
\end{theorem}

\begin{proof}

The proof is quite similar to that of Theorem~\ref{th:expTSA}, and
thus we only sketch its main steps. We split the proof in two
parts:

\begin{description}
\item{(a)}
we first show that, for every automaton $A \in {\cal B}({\cal
C}_k)$ over $\Gamma(\Sigma)$, there exists a formula $\varphi_A
\in {\rm MSO}_{{\cal P}_\Sigma}[<, (\downarrow_i)_{i=0}^{k-1}]$
over ${\cal P}_\Sigma = \{P_a \sep a \in \Sigma\}$ such that
${\cal L}(A) \cap ITS_k(\Sigma) = {\cal M}(\varphi_A)$;
\item{(b)}
then we show that, for every formula $\varphi \in {\rm MSO}_{\cal
P}[<, (\downarrow_i)_{i=0}^{k-1}]$, there exists an automaton
$A_\varphi \in {\cal B}({\cal C}_k)$ over some $\Gamma(2^{\cal
P})$ such that ${\cal M}(\varphi) = {\cal L}(A_\varphi) \cap
ITS_k({\cal P})$.
\end{description}

The embedding of automata into formulas is performed by encoding
the combined acceptance condition for ${\cal B}({\cal
C}_k)$-automata into ${\rm MSO}_{\cal P}[<,
(\downarrow_i)_{i=0}^{k-1}]$. The B\"uchi acceptance condition
have to be implemented over the leftmost path of the structure,
and the finite tree automata acceptance condition have to be
constrained to hold over almost $k$-ary trees rooted at nodes in
the leftmost path of the structure. The embedding of formulas into
automata takes advantage of the closure properties of ${\cal
B}({\cal C}_k)$-automata over UULSs.
\end{proof}

We can exploit finite tree sequence automata to provide the (full)
second-order theory of UULSs with an expressively complete
temporal logic counterpart. We know that ${\cal B}
\leftrightarrows \qltl$ and ${\cal B} \leftrightarrows \eqltl$,
and that ${\cal C}_k \leftrightarrows \qdctls$ and ${\cal C}_k
\leftrightarrows \eqdctls$. Since almost $k$-ary finite tree
automata are closed under Boolean operations,
Theorem~\ref{th:traexp} allows us to conclude that that
$\tl{\qltl}{\qdctls} \leftrightarrows {\cal B}({\cal C}_k)$ and
$\tl{\eqltl}{\eqdctls} \leftrightarrows {\cal B}({\cal C}_k)$ over
infinite sequences of almost $k$-ary finite trees. Since
increasing $k$-ary tree sequences are infinite sequences of almost
$k$-ary trees, the above equivalences hold over increasing $k$-ary
tree sequences as well. From Theorem~\ref{th:expkFTSA}, we have
that $\tl{\qltl}{\qdctls} \leftrightarrows {\rm MSO}_{\cal
P}[<_{pre}, (\downarrow_i)_{i=0}^{k-1}]$ and
$\tl{\eqltl}{\eqdctls} \leftrightarrows {\rm MSO}_{\cal
P}[<_{pre}, (\downarrow_i)_{i=0}^{k-1}]$. Such a result is
summarized by the following theorem.

\begin{theorem}
\thC{Expressiveness of $\tl{\qltl}{\qdctls}$ and
$\tl{\eqltl}{\eqdctls}$} \label{th:expQTL-UULS} \noindent
$\tl{\qltl}{\qdctls}$ and $\tl{\eqltl}{\eqdctls}$ are as
expressive as ${\rm MSO}_{\cal P}[<_{pre}, (\downarrow_i)_{i =
0}^{k-1}]$, when interpreted over UULSs.
\end{theorem}

The (nonelementary) decidability of $\tl{\qltl}{\qdctls}$ and
$\tl{\eqltl}{\eqdctls}$ immediately follows from that of ${\rm
MSO}_{\cal P}[<_{pre}, (\downarrow_i)_{i = 0}^{k - 1}]$ over
UULSs. A natural question arises at this point: is
$\tl{\eqltl}{\eqdctls}$ elementary decidable as in the case of
DULSs? In order to answer this question, we study the decidability
and complexity of the emptiness problem for finite tree sequence
automata over increasing $k$-ary tree sequences. Such a problem
can be formulated as follows: given an automaton $A \in {\cal
B}({\cal C}_k)$, is there an increasing $k$-ary tree sequence
accepted by $A$? (Equivalently, does ${\cal L}(A) \cap
ITS_k(\Sigma) \neq \emptyset$?) The (nonelementary) decidability
of such a problem immediately follows from
Theorem~\ref{th:expkFTSA}, since, given an automaton $A$, we can
build an equivalent monadic formula $\varphi_A$ and check its
satisfiability over UULSs. In the following, we give a necessary
and sufficient condition that solves the problem in {\em
elementary} time.

\medskip

Let $A = (Q,q_0,\Delta,F)$ be an automaton in ${\cal B}({\cal
C}_k)$ over the alphabet $\Gamma(\Sigma)$ (finite subset of ${\cal
C}_k$). Clearly, ${\cal L}(A) \neq \emptyset$ is necessary for
${\cal L}(A) \cap ITS_k(\Sigma) \neq \emptyset$. However, it is
not sufficient. By definition of combined acceptance condition for
$A$, we have that ${\cal L}(A) \neq \emptyset$ if and only if
there is a finite sequence $q_0, q_1, \ldots q_m$ of distinct
states in $Q$, a finite sequence $X_0, X_1, \ldots X_m$ of ${\cal
C}_k$-automata and $j \in \{0, \ldots m\}$ such that:

\begin{enumerate}
\item $\Delta(q_i,X_i,q_{i+1})$, for every $i=0, \ldots m-1$,
and $\Delta(q_m,X_m,q_j)$;
\item $q_j \in F$;
\item ${\cal L}(X_i) \neq \emptyset$, for every $i=0, \ldots m$
\end{enumerate}

To obtain a necessary and sufficient condition for ${\cal L}(A)
\cap ITS_k(\Sigma) \neq \emptyset$, we have to strengthen
condition (3) as follows. Let $T^{i}_{k}(\Sigma)$ be the set of
almost $k$-ary finite trees of height $i$:

\begin{description}
\item{3'.}
(3'a) ${\cal L}(X_i) \cap T^{i}_{k}(\Sigma) \neq \emptyset$, for
every $i=0, \ldots j-1$, and (3'b) ${\cal L}(X_i) \cap T^{i + y
\cdot l }_{k}(\Sigma) \neq \emptyset$, for every $i=j, \ldots m$
and $y \geq 0$, where $l = m-j+1$.
\end{description}

The conjunction of conditions (1,2,3') is a necessary and
sufficient condition for ${\cal L}(A) \cap ITS_k(\Sigma) \neq
\emptyset$.  We show that conditions (1,2,3') are elementarily
decidable. Clearly, there are elementarily many runs in $A$
satisfying conditions (1,2).  The following nontrivial
Lemma~\ref{lm:cond4} shows that condition 3' is elementarily
decidable.

\begin{lemma}\label{lm:cond4}
\noindent Let $X$ be a almost $k$-ary finite tree automaton, and
$a,l \geq 0$. Then, the problem ${\cal L}(X) \cap T^{a + y \cdot l
}_{k}(\Sigma) \neq \emptyset$, for every $y \geq 0$, is
elementarily decidable.
\end{lemma}

\begin{proof}

Let $X = (Q,q_0,\Delta,F)$ over $\Gamma(\Sigma)$. If $l=0$, then
the problem reduces to checking ${\cal L}(X) \cap
T^{a}_{k}(\Sigma) \neq \emptyset$, for some $a \geq 0$. For every
$a \geq 0$, the set $T^{a}_{k}$ is finite and hence regular. Since
almost $k$-ary finite tree automata are elementarily closed under
Boolean operations and elementarily decidable, we conclude that in
this case the condition is elementarily effective.

Suppose now  $l > 0$. For the sake of simplicity, we first give
the proof for finite {\em sequence} automata, and then we discuss
how to modify it to cope with the case of almost $k$-ary finite
tree automata. Hence, let $X$ be a finite sequence automaton. We
have to give an elementarily effective procedure that checks
whether $X$ recognizes at least one sequence of length $a$, at
least one of length $a+l$, at least one of length $a+2l$, and so
on. Without loss of generality, we may assume that the set of
final states of $X$ is the singleton containing $q_{fin} \in Q$.
Hence, the problem reduces to check, for every $y \geq 0$, the
existence of a path from $q_0$ to $q_{fin}$ of length $a + y \cdot
l$ in the state-transition graph associated with $X$. We thus need
to solve the following problem of Graph Theory, which we call the
{\em Periodic Path Problem} (PPP for short):

\begin{quote}
Given a finite directed graph $G = (N,E)$, two nodes $q_1,q_2 \in
N$, and two natural numbers $a,l \geq 0$, the question is: for
every $y \geq 0$, is there a path in $G$ from $q_1$ to $q_2$ of
length $a + y \cdot l$?
\end{quote}

In the following, we further reduce the PPP to a problem of Number
Theory. Let $\Pi_{q_1,q_2}(G)$ be the set of paths from $q_1$ to
$q_2$ in the graph $G$. Given $\pi \in \Pi_{q_1,q_2}(G)$, we
denote by $\pi^\circlearrowleft$ the path obtained by eliminating
cyclic subpaths from $\pi$. That is, if $\pi$ is acyclic, then
$\pi^\circlearrowleft = \pi$. Else, if $\pi = \alpha q' \beta q'
\gamma$, then $\pi^\circlearrowleft = \alpha^\circlearrowleft q'
\gamma^\circlearrowleft$. Let $\sim_{q_1,q_2}$ be the relation on
$\Pi_{q_1,q_2}(G)$ such that $\pi_1 \sim_{q_1,q_2} \pi_2$ if and
only if $\pi_{1}^{\circlearrowleft} = \pi_{2}^{\circlearrowleft}$.
Note that $\sim_{q_1,q_2}$ is an equivalence relation of finite
index. For every equivalence class $[\pi]_{\sim_{q_1,q_2}}$, we
need a formula expressing the length of a generic path in the
class. Note that every path in $[\pi]_{\sim_{q_1,q_2}}$ differs
from any other path in the same class only for the presence of
some cyclic subpaths. More precisely, let $\mu$ be the shortest
path in $[\pi]_{\sim_{q_1,q_2}}$, let $C_1, \ldots C_n$ be the
cycles intersecting $\pi$, and let $w_1, \ldots w_n$ be their
respective lengths. Note that $\mu$ does not cycle through any
$C_i$. Every path in $[\pi]_{\sim_{q_1,q_2}}$ starts from $q_1$,
cycles an arbitrary number of times (possibly zero) through every
$C_i$, and reaches $q_2$. It is easy to see that the length of an
arbitrary path $\sigma \in [\pi]_{\sim_{q_1,q_2}}$ is given by the
parametric formula: $$ |\sigma| = |\mu| + \sum_{i=1}^{n} x_i \cdot
w_i,$$ where $x_i \geq 0$ in the number of times the path $\sigma$
cycles through $C_i$.

\noindent Let $[\pi_1]_{\sim_{q_1,q_2}}, \ldots,
[\pi_m]_{\sim_{q_1,q_2}}$ be the equivalence classes of
$\sim_{q_1,q_2}$. For every $j = 1, \ldots m$, let $\mu_j$ be the
shortest path in $[\pi_j]_{\sim_{q_1,q_2}}$, let $C_{1}^{j},
\ldots C_{n}^{j}$ be the the cycles intersecting $\pi_j$, and let
$w_{1}^{j}, \ldots w_{n}^{j}$ be their respective lengths.
Moreover, let $$Y_j = \{y \geq 0 \sep \E x_{1}, \ldots x_{n} \geq
0 \, (|\mu_j| + \sum_{i=1}^{n} x_i \cdot w_{i}^{j} = a + y \cdot
l)\}.$$ The PPP reduces to the following problem of Number Theory:

\begin{quote}
Do the sets $Y_1, \ldots Y_m$ cover the natural numbers? That is,
does $\bigcup_{j=1}^{m} Y_j = \mathbb{N}$?
\end{quote}

We now solve the latter problem.  Let $w_i \geq 0$, for $i = 1,
\ldots n$. We are interested in the form of the set $S =
\{\sum_{i=1}^{n} x_i \cdot w_i \sep x_i \geq 0\}$. Let $W = (w_1,
\ldots w_n)$ and let $d = GCD(W)$ (the greatest common divisor of
$\{w_1, \ldots, w_n\}$). We distinguish the cases $d = 1$ and $d
\neq 1$. If $d = 1$, then it is easy to see that: $$S = E \cup \{j
\sep j \geq k\},$$ where $E$ is a finite set of {\em exceptions}
such that $max(E) < k$, and $k = (w_r-1) \cdot (w_s-1)$, with $w_r
= min(W)$ (the minimum of $\{w_1, \ldots w_n\}$) and $w_s = min(W
\setminus w_{r})$. If $d \neq 1$, then consider the set $S' =
\{\sum_{i=1}^{n} x_i \cdot w_i/d \sep x_i \geq 0\}$. Clearly,
$GCD(w_1/d, \ldots w_n/d) = 1$ and hence, as above, $S' = E' \cup
\{j \sep j \geq k'\}$ for some finite set $E'$ and some $k' \in
\mathbb{N}$. Therefore, in this case, $$S = E' \cdot d \cup \{j
\sep j \geq k' \cdot d \And d \ \tt{DIV} \ j\},$$ where $d \
\tt{DIV} \ j$ means that $d$ is a divisor of $j$.

Summing up, in any case, the set $S$ can be described as follows:
$$S = E \cup \{k + j \cdot d \sep j \in \mathbb{N}\},$$ for some
finite (computable) set $E$, some (computable) $k \in \mathbb{N}$,
and $d = GCD(W)$. In other words, the set $S$ is the union of a
finite and computable set of exceptions and an arithmetic
progression.

Now we consider the equation $\sum_{i=1}^{n} x_{i} \cdot w_i = y
\cdot l$. Our aim is to describe the set $Y = \{y \geq 0 \sep \E
x_{1}, \ldots x_{n} \geq 0 \, (\sum_{i=1}^{n} x_{i} \cdot w_i = y
\cdot l)\}$ in a similar way. Let $e = GCD(d,l)$, $l = l' \cdot e$
and $d = d' \cdot e$. We have that:

\medskip

$\begin{array}{lll} y \in Y & \mbox{ iff } & \\ y \cdot l \in S &
\mbox{ iff } & \\ y \cdot l \in E \Or y \cdot l \geq k \And d \
\tt{DIV} \ y \cdot l & \mbox{ iff } & \\ y \cdot l \in E \Or y
\geq \lceil k/l \rceil \And d' \cdot e \ \tt{DIV} \ y \cdot l'
\cdot e  & \mbox{ iff } & \\ y \cdot l \in E \Or  y \geq \lceil
k/l \rceil \And d' \ \tt{DIV} \ y
\end{array}$

\medskip

Therefore, the set $Y$ is the union of a finite and computable set
and an arithmetic progression, i.e., $$Y = E' \cup \{k' + j \cdot
d' \sep j \in \mathbb{N}\},$$ for some finite (computable) set
$E'$, some (computable) $k' \in \mathbb{N}$, and $d' = d / GCD(d,$
$l)$. The set $Y = \{y \geq 0 \sep \E x_{1}, \ldots x_{n} \geq 0
\, (\sum_{i=1}^{n} x_{i} \cdot w_i = a + y \cdot l)\}$, with $a
\in \mathbb{N}$, can be described in the same way.

We have shown that, for $i = 1, \ldots, m$, every $Y_i$ has the
form $E_i \cup \{k_i + y \cdot d_i \sep y \geq 0\}$ for some
finite $E_i$, and some $k_i,d_i \in \mathbb{N}$. We now give a
solution to the problem $\bigcup_{i=1}^{m} Y_i = \mathbb{N}$. Let
$k_r = min \{k_1, \ldots, k_m\}$ and $D = LCM(d_1, \ldots, d_m)$
(the least common multiple of $\{d_1, \ldots, d_m\}$). The
algorithm works as follows: for every $k < k_r$, we check whether
$k \in Y_i$ for some $i = 1, \ldots, m$. If this is not the case,
the problem has no solution. Otherwise, we verify whether, for
every $j = 0, \ldots, D-1$, $k_r + j \in Y_i$ for some $i = 1,
\ldots, m$. If this is the case, then we have a solution,
otherwise, there is no solution. Note that a solution can be
described in terms of an ultimately periodic word $w = uv^\omega$,
with $u,v \in \{1, \ldots m\}^*$, such that, for every $i \geq 0$,
$w(i) = j$ means that a path from $q_1$ to $q_2$ in the graph $G$
belongs to the $j$-th equivalence class
$[\pi_j]_{\sim_{q_1,q_2}}$.

The above algorithm solves the periodic path problem in doubly
exponential time in the number $n$ of nodes of the graph $G$. The
number of equivalence classes of the relation $\sim_{q_{1},q_{2}}$
over the set of paths from $q_{1}$ to $q_{2}$ in $G$ may be
exponential in $n$. Thus, we have $m$ sets $Y_1, \ldots, Y_m$,
each one associated with a relevant equivalence class, and $m =
{\cal O}(2^n)$. Every set $Y_i$ can be represented in polynomial
time as $E_i \cup \{k_i + y \cdot d_i \sep y \geq 0\}$ for some
finite $E_i$, and some $k_i,d_i \in \mathbb{N}$. Note that the
cardinality of $E_i$ is bounded by $k_i$, $k_i = {\cal O}(n^2)$,
and $d_i = {\cal O}(n)$. The final step of the procedure makes
$k_0 + D$ membership tests with respect to some set $Y_i$, where
$k_0 = min \{d_1, \ldots d_m\}$, and $D =  LCM(d_1, \dots d_m)$.
Each test is performed in ${\cal O}(1)$. Moreover, $D$ is bounded
by ${d_{0}}^{m}$, where $d_0 = max \{d_1, \ldots d_m\}$, and hence
$D = {\cal O}(2^{2^n})$. Hence, the procedure works in doubly
exponential time.

The general case of finite trees is similar. Let $X$ be a finite
almost $k$-ary tree automaton. A path from $q_1$ to $q_2$
corresponds to a run of $X$ such that the run tree is complete and
$k$-ary, the root of the run tree is labeled with state $q_1$ and
the leaves of the run tree are labeled with state $q_2$. A cycle
is a path from $q$ to $q$. The problem is to find, for every $y
\geq 0$, a path from the initial state $q_0$ to the final state
$q_{fin}$ of length $a + y \cdot l$. The rest of the proof follows
the same reasoning path of the proof for sequence automata.
\end{proof}

From Lemma \ref{lm:cond4}, it follows that, given a ${\cal
B}({\cal C}_k)$-automaton $A$, we have an algorithm to solve the
problem ${\cal L}(A) \cap ITS_k(\Sigma) \neq \emptyset$ in doubly
exponential time in the size of $A$.

\begin{theorem}
\label{th:deckFTSA} \noindent The emptiness problem for finite
tree sequence automata over UULSs is in 2EXPTIME.
\end{theorem}

Since $\tl{\eqltl}{\eqdctls}$ formulas can be elementarily
converted into ${\cal B}({\cal C}_k)$ automata, we have the
desired result.

\begin{theorem}
\label{th:compTL-UULS} \thC{Complexity of $\tl{\eqltl}{\eqdctls}$}
\noindent The satisfiability problem for $\tl{\eqltl}{\eqdctls}$
over UULSs is in ELEMENTARY.
\end{theorem}

We conclude the section by giving some examples of meaningful
timing properties that can be expressed in (fragments of)
$\tl{\eqltl}{\eqdctls}$ interpreted over UULSs. As a first
example, consider the property `$P$ holds at every point of the
finest layer $T^0$ whose distance from the origin of the layer
$0_0$ is a power of two ($1_0, 2_0, 4_0, 8_0$, and so on)' over a
binary UULS. Such a property can be expressed in
$\tl{\ptl}{\dctls}$ as follows: $$\Next \Box \, \op{E X_1}
\op{G}((\op{X} {\tt true} \Imp \op{X_0} {\tt true}) \And (\neg
\op{X} {\tt true} \Imp P)).$$ Notice that the property `$P$ holds
on every point $2^i$, with $i \in \mathbb{N}$' cannot be expressed
in $\qltl$. As a second example, the property `$P$ holds on every
even point of the leftmost path' can be expressed in
$\tl{\eqltl}{\dctls}$ as follows: $$\E Q(Q \And \Next \neg Q \And
\Box(Q \IImp \Next \Next Q) \And \Box(Q \Imp P)).$$ As already
pointed out, this property cannot be expressed in
$\tl{\ptl}{\dctls}$, since $\ptl$ cannot express the property `$P$
holds on every even point'~\cite{W83}.

As in the case of DULSs, there are some natural properties of
UULSs that cannot be easily captured in $\tl{\eqltl}{\eqdctls}$.
As an example, it is not easy to express the property `$P$ holds
on every even point of the finest domain $T^0$'.

\section{The specification of a high voltage station}
\label{sec:exa}

In this section, we exemplify the concrete use of temporalized
logics as specification formalisms by providing (an excerpt of)
the specification of a supervisor that automates the activities of
a High Voltage (HV) station devoted to the end user distribution
of energy generated by power plants \cite{M96}. We first show how
relevant timing properties of such a system can be expressed in
monadic second-order languages, and then we give their simpler
temporalized logic formulations.

Each HV station is composed of bays, connecting generation units
to the distribution line. A bay consists of circuit breakers and
insulators. They are both switches, but an expensive circuit
breaker can interrupt current in a very short time (50 millisecond
or even less), while a cheap insulator is not able to interrupt a
flowing current and it has a switching time of a few seconds. Let
us consider a simple HV station consisting of two bars {\tt b1}
and {\tt b2} connected to different power units, a distribution
line {\tt l}, and two bays {\tt pb} (parallel bay) and {\tt lb}
(line bay). The parallel bay shorts circuit between the two bars
{\tt b1} and {\tt b2}. It consists of two insulators {\tt ip1} and
{\tt ip2}, and one circuit breaker {\tt cbp}. It is in the state
{\tt closed} if all its switches are closed; otherwise it is {\tt
open}. The line bay connects the distribution line with either the
first bar or the second one. It consists of three insulators {\tt
ilb1}, {\tt ilb2}, and {\tt il1}, and one circuit breaker {\tt
cbl}. It is in the state {\tt closed\_on\_b1} if {\tt ilb1}, {\tt
cbl},  and {\tt il1} are closed, and in the state {\tt
closed\_on\_b2} if {\tt ilb2}, {\tt cbl},  and {\tt il1} are
closed.

We focus on the specification of the change of the bar connected
to the line from {\tt b1} to {\tt b2}. The supervisor starts its
operation by closing the parallel bay, an action that takes about
$10$ \emph{seconds}; then, it first closes the insulator {\tt
ilb2}, an action that takes about $5$ \emph{seconds} and then it
opens the insulator {\tt ilb1}, and action that takes $5$
\emph{seconds} as well; finally, it opens the parallel bay, an
action that takes other $10$ \emph{seconds}.
To model the behavior of the system, we use the predicates {\tt
change\_b1\_b2}, {\tt change\_b2\_b1}, {\tt close\_pb}, {\tt
open\_pb}, {\tt close\_ilb1}, {\tt open\_ilb1}, {\tt close\_ilb2},
and so on to denote the corresponding commands sent by the
supervisor to the various devices. Furthermore, for every system
action we identify the time granularity with respect to which it
can be considered as an instantaneous action. The change of the
bar takes about $30$ seconds, opening and closing the parallel bay
$10$ seconds, switching insulators $5$ seconds, and switching
circuit breakers $50$ milliseconds. Accordingly, we assume a
$4$-layered structure whose $4$ layers correspond to the $4$
involved time granularities, namely, {\tt 30secs}, {\tt 10secs},
{\tt 5secs}, and {\tt 50millisecs} (in \cite{FM03} we show how to
tailor temporal logics for time granularity over downward
unbounded layered structures to deal with $n$-layered structures).

In the monadic second-order language, the change of the bar is
described by the following formula, which specifies the sequence
of actions taken by the supervisor:

$$\begin{array}{ll}
  \A x.\ (T^{30secs}(x) \And change\_b1\_b2(x) \Imp &
  \E y_1.\ {\da_0}(x) = y_1 \And close\_pb(y_1) \And \\
  & \E y_2.\ {+1}_{10secs}(y_1,y_2) \And \\
  & \E y_3.\ {\da_0}(y_2) = y_3 \And close\_ilb2(y_3) \And \\
  & \E y_4.\ {+1}_{5secs}(y_3,y_4) \And open\_ilb1(y_4) \And \\
  & \E y_5.\ {+1}_{5secs}(y_4,y_5) \And \\
  & \E y_6.\ {\da_0}(y_6) = y_5 \And open\_pb(y_6)),
\end{array}$$

\noindent where the definable predicate ${+1}_{g}(x,y)$ states
that both $x$ and $y$ belong to the layer $g$ and $y$ is the
successor $x$ with respect to $g$. Such a condition can be
expressed in temporalized logic in a much more compact and
readable way:

$$\begin{array}{ll} \op{G}(change\_b1\_b2 \Imp & \op{EX_0}
close\_pb \And \op{EX_1X_0}close\_ilb2  \And \\ & \op{EX_1X_1}
open\_ilb1 \And \op{EX_2} open\_pb)
\end{array}$$

As for the compound operation {\tt close\_pb}, let us assume that
the supervisor starts in parallel the closure of the circuit
breaker, which is completed in 50 milliseconds, and of the first
insulator, that takes about 5 seconds; then, once the first
insulator is closed, it closes the second one. Such an operation
can be specified by the following classical formula:

$$\begin{array}{ll}
  \A x.\ (T^{10secs}(x) \And close\_pb(x) \Imp &
  \E y_1.\ {\da_0}(x) = y_1 \And close\_ip1(y_1) \And \\
  & \E y_2.\ {\da_0}(y_1,y_2) \And close\_cbp(y_2) \And \\
  & \E y_3.\ {+1}_{5secs}(y_1,y_3) \And close\_ip2(y_3),
\end{array}$$

while its temporalized version is structured as follows:

$$\begin{array}{ll} \op{G}(&(\op{EX_0} close\_pb \Imp
\op{EX_0}(\op{EX_0}(close\_ip1 \And \op{X_0} close\_cpb) \And
\op{EX_1} close\_ip2)) \And \\ & (\op{EX_1} close\_pb \Imp
\op{EX_1}(\op{EX_0}(close\_ip1 \And \op{X_0} close\_cpb) \And
\op{EX_1} close\_ip2)) \And \\ & (\op{EX_2} close\_pb \Imp
\op{EX_2}(\op{EX_0}(close\_ip1 \And \op{X_0} close\_cpb) \And
\op{EX_1} close\_ip2))).
\end{array}$$

\section{Conclusions and future work}
\label{sec:cfw}

In this paper, we provided the monadic second-order theories of
DULSs and UULSs with expressively complete and elementarily
decidable temporal logic counterparts. To this end, we defined
temporalized automata, which can be seen as the
automaton-theoretic counterpart of temporalized logics, and showed
that relevant properties, such as closure under Boolean
operations, decidability, and expressive equivalence with respect
to temporal logics, transfer from component automata to
temporalized ones. Then, we exploited temporalized automata to
successfully solve the problem of finding the temporal logic
counterparts of the given theories of time granularity.

As a matter of fact, some forms of automaton combination, which
differ from temporalization in various respects, have been
proposed in the literature to increase the expressive power of
temporal logics. As an example, extensions of PLTL with
connectives defined by means of finite automata over
$\omega$-strings are investigated in \cite{VW94}. To gain the
expressive power of the full monadic second-order theory of
$(\omega,<)$, Vardi and Wolper's Extended Temporal Logic (ETL)
replaces the until operator of PLTL by an infinite bunch of
automata connectives, that is, ETL allows formulas to occur as
arguments of an automaton connective (as many formulas as the
symbols of the automaton alphabet are). Given the well-known
correspondence between formulas and automata, the application of
automata connectives to formulas can be viewed as a form of
automata combination. An extension of $\ctls$ that substitutes ETL
operators for PLTL ones is given in \cite{Da94}. However, the
switch from PLTL to ETL does not involve any change in the domain
of interpretation ($\omega$-structures in the first case, binary
trees in the latter). On the contrary, in the case of temporalized
automata/logics, component automata/temporal logics refer to
different temporal structures, and thus their combination is
paired with a combination of the underlying temporal structures.

We are developing our research on temporalized logics and automata
for time granularity in various directions. First of all, we are
trying to improve the complexity bound for the satisfiability
problem for $\tl{\eqltl}{\eqdctls}$ over UULSs. Second, we are
investigating the relationships between temporalized and classical
automata. On the one hand, the languages recognized by
temporalized automata are structurally different from those
recognized by classical automata, e.g., {\tt B\"uchi(B\"uchi)}
automata recognize infinite strings of infinite strings. On the
other hand, this fact does not imply that language problems for
temporalized automata cannot be reduced to the corresponding
problems  for classical automata. As an example, the emptiness
problem for {\tt B\"uchi(B\"uchi)} automata can actually be
reduced to the emptiness problem for {\tt B\"uchi} automata. We
are exploring the possibility of defining similar reductions for
more complex temporalized automata. Finally, we are exploring the
possibility of extending our correspondence results to other forms
of logic combination, such as independent combination and join
\cite{GKWZ03}.

\section*{Acknowledgements}
We would like to thank the anonymous reviewers for their comments
and criticisms, that helped us to improve the paper, as well as
Johan van Benthem, Ahmed Bouajjani, Marcelo Finger, Valentin
Goranko, Maarten de Rijke, and Wolfgang Thomas for their positive
feedback on the work reported in the paper. Thanks also to Pietro
Corvaia, Alberto Policriti, and Franca Rinaldi for the useful
discussions about the proof of Lemma \ref{lm:cond4}.

\bibliographystyle{acmtrans}

\end{document}